\documentclass{pasj01}
\Received{$\langle$2020 June 1$\rangle$}
\Accepted{$\langle$2020 December 1$\rangle$}
\Published{$\langle$publication date$\rangle$}
\begin{document}

\title{Observations of the [CI] ($^3P_1$--$^3P_0$) emission toward the massive star-forming region RCW38: further evidence for highly-clumped density distribution of the molecular gas}
\author{Natsuko Izumi,$^{1,2,3,*}$ Yasuo Fukui,$^{4}$ Kengo Tachihara,$^{4}$ Shinji Fujita,$^{5}$ Kazufumi Torii,$^{6}$ Takeshi Kamazaki,$^{2}$
Hiroyuki Kaneko, $^{7,2}$ Andrea Silva,$^{2}$ Daisuke Iono,$^{2}$ Munetake Momose,$^{1}$ Kanako Sugimoto,$^{2}$ Takeshi Nakazato,$^{2}$ George Kosugi,$^{2}$ Jun Maekawa,$^{2}$ Shigeru Takahashi,$^{6}$
Akira Yoshino,$^{2}$ and Shin'ichiro Asayama$^{2}$}%
\altaffiltext{1}{Colledge of Science, Ibaraki University, 2-1-1 Bunkyo, Mito, Ibaraki 310-8512, Japan}
\altaffiltext{2}{National Astronomical Observatory of Japan, National Institutes of Natural Sciences, 2-21-1 Osawa, Mitaka, Tokyo 181-8588, Japan}
\altaffiltext{3}{Institute of Astronomy and Astrophysics, Academia Sinica, No. 1, Section 4, Roosevelt Road, Taipei 10617, Taiwan}
\altaffiltext{4}{Department of Physics, Nagoya University, Chikusa-ku, Nagoya, Aichi 464-8601, Japan}
\altaffiltext{5}{Department of Physical Science, Graduate School of Science, Osaka Prefecture University, 1-1 Gakuen-cho, Naka-ku, Sakai, Osaka 599-8531, Japan}
\altaffiltext{6}{Nobeyama Radio Observatory, National Astronomical Observatory of Japan (NAOJ), National Institutes of Natural Sciences (NINS), 462-2 Nobeyama, Minamimaki, Minamisaku-gun, Nagano 384-1305, Japan}
\altaffiltext{7}{ Graduate School of Education, Joetsu University of Education, 1 Yamayashiki-machi, Joetsu, Niigata, 943-8512, Japan}
\email{nizumi923@gmail.com}

 
\maketitle

\begin{abstract}
We present observations of the $^3P_1$--$^3P_0$ fine-structure line of atomic carbon using the ASTE 10 m sub-mm telescope towards RCW38,
the youngest super star cluster in the Milky Way.
The detected [CI] emission is compared with the CO $J$ = 1--0 image cube presented in \citet{Fukui2016} which has an angular resolution of 40$^{\prime \prime}$ ($\sim$ 0.33 pc).
The overall distribution of the [CI] emission in this cluster is similar to that of the $^{13}$CO emission.
The optical depth of the [CI] emission was found to be $\tau$ = 0.1--0.6, suggesting mostly optically thin emission. 
An empirical conversion factor from the [CI] integrated intensity to the H$_2$ column density was estimated as
$X_{\rm [CI]}$ = 6.3 $\times$ 10$^{20}$ cm$^{-2}$ K$^{-1}$ km$^{-1}$ s (for visual extinction: $A_V$ $\le$ 10 mag)
and 1.4 $\times$ 10$^{21}$ cm$^{-2}$ K$^{-1}$ km$^{-1}$ s (for $A_V$ of 10--100 mag).
The column density ratio of the [CI] to CO ($N_{\rm [CI]}/N_{\rm CO}$) was derived as $\sim$ 0.1 for $A_V$ of 10--100 mag,
which is consistent with that of the Orion cloud presented in \citet{Ikeda2002}.
However, our results cover an $A_V$ regime of up to 100 mag, which is wider than the coverage found in Orion, which reach up to $\sim$ 60 mag.
Such a high [CI]/CO ratio in a high $A_V$ region is difficult to be explained by the plane-parallel photodissociation region (PDR) model,
which predicts that this ratio is close to 0 due to the heavy shielding of the ultraviolet (UV) radiation.
Our results suggest that the molecular gas in this cluster is highly clumpy, allowing deep penetration of UV radiation even at averaged $A_V$ values of 100 mag.
Recent theoretical works have presented models consistent with such clumped gas distribution with a sub-pc clump size (e.g., \cite{Tachihara2018}).
\end{abstract}

\section{Introduction}\label{sec:intro}
Interstellar molecular clouds are the site of star formation in galaxies.
Ultraviolet (UV) photons emitted from OB-stars have significantly influence the structure and chemical diversity of the molecular clouds,
and the behavior of ionized regions within them has been of great interest in the last decades (e.g., \cite{Hollenbach1997}).
A partially ionized layer in a molecular cloud faced to UV photons is known as a ``photodissociation region (PDR)'' and its structure has been intensively studied theoretically (e.g., \cite{Tielens1985,Hollenbach1991}).
The plane-parallel model (e.g., \cite{Hollenbach1999}) of a PDR suggests that these regions comprise layers with different chemical abundances such as, H/H$_2$ and C$^{+}$/C/CO,
in the order of the ionization/dissociation energy where relatively uniform density distribution was assumed.
This model predicts the concentration of atomic carbon in a visual extinction ($A_V$) range of 3--6 mag (e.g., \cite{Tielens1985,vanDishoeck1988,Hollenbach1991}).
The PDR theory must be observationally tested to understand the physics and chemistry in PDRs better.
In particular, atomic carbon lies in the intermediate layer next to CO, and its transitions can provide for a sensitive probe of the carbon chemistry in PDRs.
After the pioneering studies (e.g., Phillips et al. 1980; Keene et al. 1985),
the $^3P_1$--$^3P_0$ fine-structure line of atomic carbon [CI] (492.16967 GHz) has been known to be a good tracer of interstellar atomic carbon.
Past observations have revealed that [CI] coexists with CO and their intensities correlate with each other.
Large-area [CI] mapping observations toward the Galactic plane with low-UV ($G_0$ $\sim$ 1.7--5.1: typical average interstellar value) have shown that the distribution of the [CI] emission
is almost similar to that of the CO emission (e.g., \cite{Oka2005,Burton2015}).
[CI] observations in regions with intense UV radiation also show that the [CI] and CO distributions of Orion and Carina clouds  are similar to each other,
where $G_0$ is 10$^4$--10$^5$ (e.g. \cite{Ikeda2002,Kramer2008,Shimajiri2013}).
 \citet{Shimajiri2013} performed [CI] mapping observations at a resolution of 21$\farcs$3 ($\sim$ 0.04 pc) toward the Orion A cloud
with the Atacama Submillimeter Telescope Experiment (ASTE) 10 m telescope.
They found that [CI] shows a similar distribution that $^{13}$CO and did not find a layer of [CI] outside the $^{13}$CO distribution.
These observations indicate that the [CI] distribution is similar to CO distribution, suggesting that the plane-parallel PDR model presented in the early works may not be appropriate for molecular clouds.
Therefore, this model must be tested by analyzing the behavior of [CI] in more PDR samples. 

RCW38 \citep{Rodgers1960} is an HII region of intense UV radiation located at a relatively small distance of 1.7 kpc from the Sun.
It contains $\sim$ 20 candidate O-stars, slightly larger than in the Orion KL region,
in a volume of less than 1 pc \citep{Wolk2006} and is very luminous with $L$ $\sim$ 7 $\times$ 10$^5$ $L_\odot$ \citep{Furniss1975}.
This source is, therefore, ideal to test the PDR model.
There are two remarkable infrared peaks in the central part of RCW38 \citep{Frogel1974}.
The brightest 2.2 $\micron$ source is labeled as IRS 2, and corresponds to a O5.5 binary system \citep{DeRose2009}.
The brightest 10 $\micron$ feature located 0.1 pc west of IRS 2 is labeled as IRS 1.
This is a dust ridge extending for 0.1--0.2 pc in the north--south direction \citep{Smith1999}. 
Furthermore, RCW38 is the youngest super star cluster in the Galaxy ($\lesssim$ 1 Myr;  e.g., \cite{Wolk2006,Wolk2008,Winston2011}).

Molecular clouds were observed by \citet{Zinchenko1995} in the central region of RCW38 in the CS emission (resolution: $\sim$ 1$^{\prime}$) to resolve the ring-like shape surrounding the O-star candidates.
\citet{Yamaguchi1999} observed a large area of RCW38 in the $^{13}$CO(1-0) emission with the NANTEN 4 m telescope at a lower resolution (2$^{\prime}$.7) and found an extended molecular cloud
surrounding the HII region.
\citet{Gyulbudaghian2008} performed $^{12}$CO(1-0) observation with the SEST telescope at 45$^{\prime \prime}$ resolution for a small 160$^{\prime \prime}$ $\times$ 160$^{\prime \prime}$ region.
The $^{12}$CO(1-0) emission shows two different velocity components at $-$3--2 km s$^{-1}$ and 3--8 km s$^{-1}$.
Recently, \citet{Fukui2016} conducted observations of $^{12}$CO(1-0), $^{13}$CO(1-0), C$^{18}$O(1-0), and $^{12}$CO(3-2) toward RCW38 with Mopra, ASTE, and NANTEN2 telescopes.
The observations revealed detailed distributions of two molecular clouds at the velocities of 2 and 14 km s$^{-1}$ toward RCW38.
The Ring cloud (the 2 km s$^{-1}$ cloud) exhibits a ring-like shape with a cavity ionized by the cluster and has a high H$_2$ column density of $\sim$ 10$^{23}$ cm$^{-2}$,
whereas, the Finger cloud (the 14 km s$^{-1}$ cloud) has a tip toward the cluster and has a lower H$_2$ column density of $\sim$ 10$^{22}$ cm$^{-2}$. 
They suggest that the clouds are incidentally colliding with each other to trigger the formation of the $\sim$ 20 O-star candidates, which are located within $\sim$ 0.5 pc of the cluster center in the Ring cloud.
The collision is evidenced by the bridge feature connecting the two clouds, which is localized toward the O-star candidates.
The collision timescale is estimated to be $\sim$ 0.1 Myrs from the cloud size and velocity separation, corresponding the fact that the RCW38 cluster is very young.

In this paper, we report the results of [CI] observations toward RCW38 with ASTE.
The wide-field (6\farcm0 $\times$ 8\farcm0 $\sim$ 3 pc $\times$ 4 pc) and high-resolution (17$^{\prime \prime}$ $\sim$ 0.14 pc)
allowed us to reveal the detailed [CI] gas distributions and physical conditions, such as optical depth and column density.
Combining these with the results of past CO observations by \citet{Fukui2016}, we investigate the relationship between CO and [CI] emissions around the PDR and molecular clouds in RCW38.
Section~2 describes the basic observational setup and data reduction.
Section~3 presents the distribution and physical properties of the CO and [CI] emissions.
In Section~4, we discuss the relationship between CO and [CI] emissions in RCW38 and investigate the PDR model.
We conclude the paper in Section~5.

\section{Observation and data reduction}\label{sec:obs_data}
\subsection{Observation}\label{sec:obs}
We observed RCW38 in the  [CI] ($^3P_1-^3P_0$; 492.1607 GHz) line with the  ASTE 10 m telescope \citep{Ezawa2004,Kohno2004}
located at Pampa la Bola at an altitude of 4800 m.
These observations were remotely conducted on November 26, 2016 from Japan using the remote control system N-COSMOS3 \citep{Kamazaki2005}.
We used ALMA Band 8 receiver, which is a dual-polarization side-band separating (2SB) mixer receiver operating at 400--500 GHz \citep{Ito2018}.
As a back end,  we used the MAC spectrometer, which is a subset of AC45 spectrometer for the Nobeyama 45-m telescope \citep{Sorai2000}.
We configured the spectrometer with the mode of 1024-channel power spectrum of 512 MHz bandwidth, which provides a frequency resolution of 0.5 MHz, corresponding to 0.3 km s$^{-1}$ at the [CI] frequency.
The telescope half-power beam width (HPBW) is 17$^{\prime \prime}$.
We performed on-the-fly (OTF) mapping observations over a 6\farcm0 $\times$ 8\farcm0 area \citep{Sawada2008},
referencing to an off-source position at R.A. = 08$^{\rm h}$55$^{\rm m}$01\fs51 and decl. = -47$^\circ$23$^\prime$16\farcs2 (J2000).
The off-source position was free from $^{12}$CO line emission in the wide-field CO mapping data by NANTEN2 telescope presented in \citet{Fukui2016}.
The OTF mapping was performed along each of two orthogonal directions, R.A. and decl.
The intervals between the adjacent scans were 6\farcs0, which was approximately 1/3 of the HPBW.
 The data was taken every 0.1 second while
 the scans were taken at a constant slewing rate of 20$^{\prime \prime}$ sec$^{-1}$, which yielded a spatial sampling interval of 2\farcs0.
The telescope pointing was checked every two hours by five-point scans of pointing sources: O-Cet [R.A. = 02$^{\rm h}$19$^{\rm m}$20\fs79, decl. = -02$^\circ$58$^\prime$41\farcs91(J2000)]
and IRC+10216 [R.A. = 09$^{\rm h}$47$^{\rm m}$57\fs41, decl. = +13$^\circ$16$^\prime$43\farcs60 (J2000)],
and the pointing accuracy was better than $\sim$ 2$^{\prime \prime}$.
The standard chopper-wheel method was used to convert the output signals into antenna temperatures ($T^*_{\rm A}$), corrected for the atmospheric attenuation.
The typical system noise temperature was 800--1400 K in the target directions.
These observations were performed as a part of the ASTE demo science project.

\subsection{Data reduction}\label{sec:data}
The data were processed using the Common Astronomy Software Applications (CASA version 5.1.1: \cite{McMullin2007}).
After subtracting the linear baselines from the OTF spectra in the frequency domain, they were convolved with a GJINC (Gaussian $\times$ Jinc) function \citep{Mangum2007}, resampled onto a 6$^{\prime \prime}$ grid,
and summed over each 0.6 km s$^{-1}$ velocity bin,
resulting in an effective spatial resolution of $\sim$ 20$^{\prime \prime}$ in HPBW.
The adopted main beam efficiency to convert $T^*_{\rm A}$ to main-beam temperature ($T_{\rm MB}$) is 0.45.
Our target line presents bright and spatially extended structures over the observed region,
which may additionally contribute to the efficiency due to the error beam of ASTE.
Such contribution was estimated to be $<$ 20 \% based on the past measurement on Jupiter (private communication).
The systematic error of the [CI] amplitude was expected to be 7 \% in terms of $T^*_{\rm A}$,
which was dominated by the variations of the atmospheric conditions during the observations.
Finally, to remove the striping pattern generated along the scan directions,
we combined the two maps generated by scanning the R.A. and decl. directions using the basket-weaving method \citep{Emerson1988}.
The typical rms-noise level in the final cube data is 0.7 K in $T_{\rm MB}$ (corresponding to 0.3~K in $T^*_{\rm A}$) with a velocity resolution of 0.6 km s$^{-1}$.
The raw data, final cube data, and our detailed analysis procedures are available from the  ASTE web page\footnote{https://sites.google.com/site/asteobservation/home/reduction\#TOC-DEMO-SCIENCE-DATA-REDUCTION-GUIDE}.

\subsection{Previous observations}\label{sec:pre_obs}
In addition to our [CI] observation, we used the CO data from the previous observations presented in \citet{Fukui2016}: $^{12}$CO(1-0), $^{13}$CO(1-0), and C$^{18}$O(1-0) observations with
the 22m ATNF  (Australia Telescope National Facility) Mopra millimeter telescope in Australia
and $^{12}$CO(3-2) observations with the ASTE.
The angular resolution and velocity resolution of the Mopra observations were 33$^{\prime \prime}$ and 0.088 km s$^{-1}$, respectively,
and those of the ASTE observations were 22$^{\prime \prime}$ and 0.11 km s$^{-1}$, respectively.
The Mopra and ASTE data were smoothed to an HPBW of 40$^{\prime \prime}$ with a 2D Gaussian function and to 0.6 km s$^{-1}$ velocity resolution \citep{Fukui2016}.
The typical rms-noise levels in $^{12}$CO(1-0), $^{12}$CO(3-2), $^{13}$CO(1-0), and C$^{18}$O(1-0) data are 0.4 K, 0.2K, 0.2 K, and 0.1 K, respectively, in $T_{\rm MB}$ with a velocity resolution of 0.6 km s$^{-1}$.
To compare the [CI] data with the CO data and improve signal-to-noise ratios,
the ASTE [CI] data were smoothed to an HPBW of 40$^{\prime \prime}$ with a 2D Gaussian function.
The typical rms-noise level of the smoothed ASTE [CI] data is 0.3 K in $T_{\rm MB}$ with a velocity resolution of 0.6 km s$^{-1}$.

\section{Results}\label{sec:res}
\subsection{[CI] distribution of RCW38}\label{sec:res_dist}
\begin{figure*}
 \begin{center}
  \includegraphics[width=16cm]{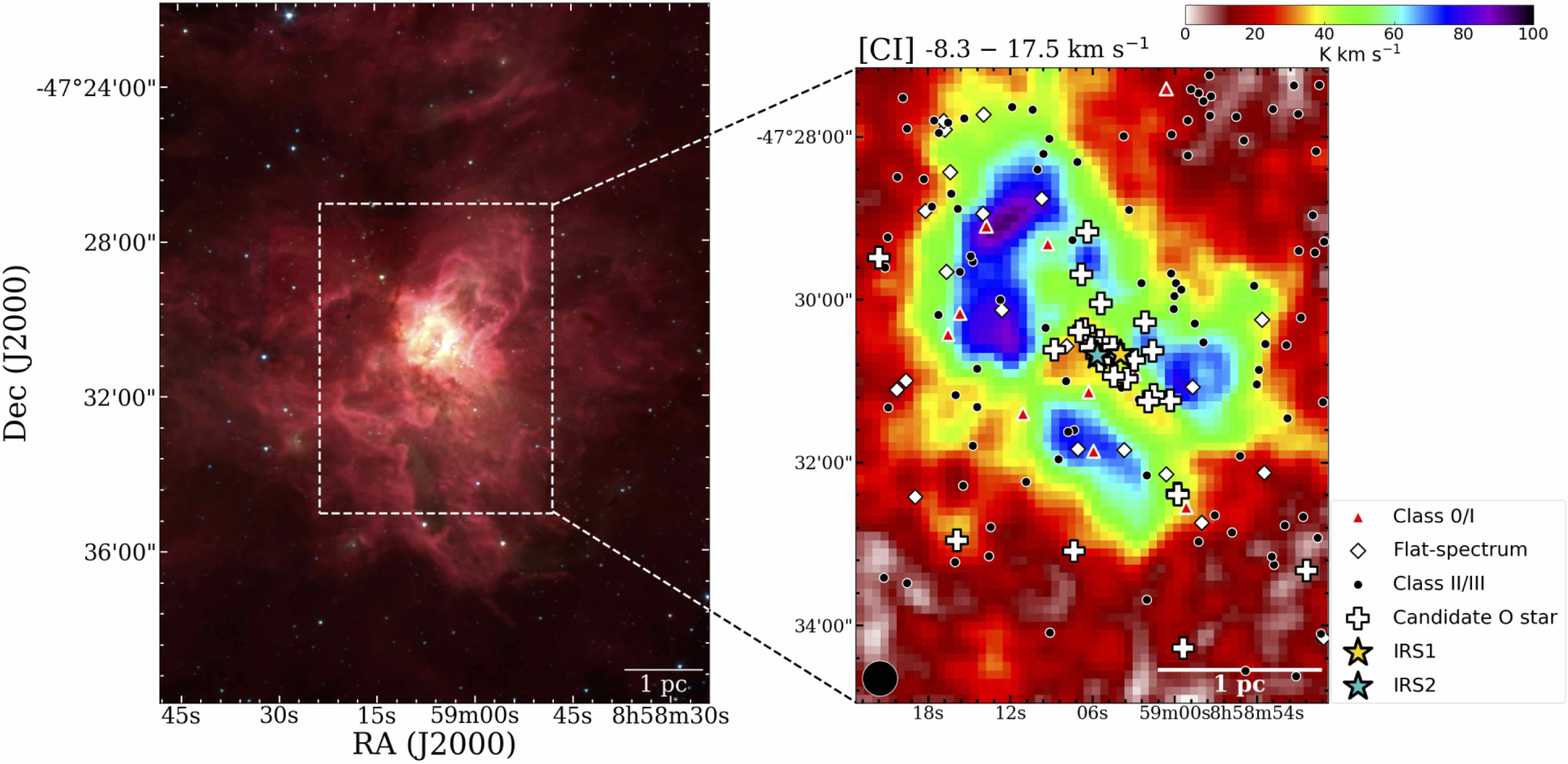} 
 \end{center}
\caption{
Mid-infrared pseudo color image of RCW38 (Left: \cite{Winston2011}) and [CI] integrated intensity distributions of RCW38 (Right).
The color image was produced by combining the 3.6, 4.5, and 8.6 $\micron$ images from {\it Spitzer} data.
The white dashed box in the left panel shows the mapping size of the ASTE observation (6$^{\prime}$ $\times$ 8$^{\prime}$).
YSOs and O star candidates obtained by \citet{Wolk2006} and \citet{Winston2011} are plotted in the right panel.
Red triangles and white diamonds indicate the class 0/I and flat-spectrum YSOs, respectively.
Black circles indicate the class II and class III YSOs \citep{Winston2011}. White crosses indicate the O-star candidates.
Yellow and cyan star symbols indicate the peaks of IRS 1 and IRS 2, respectively \citep{Wolk2006}.
These data were smoothed to an HPBW of 26$^{\prime \prime}$ with a 2D Gaussian function, and the rms noise level was reduced to a typical value of 0.5 K in $T_{\rm MB}$.
The black filled circle at the lower left corner in the right panel shows the HPBW size.}
\label{IR_FoV}
\end{figure*}

The right panel in Figure \ref{IR_FoV} shows the [CI] integrated intensity distribution, and YSO and O-star candidates \citep{Wolk2006, Winston2011} in RCW38. 
The [CI] emission has a ring-like structure with a diameter of 1--2 pc.
The O-star candidates, including IRS 2, are concentrated in the cavity. 
Several filamentary and bubble-like structures extend radially outside the ring-like structure by 1--2 pc.
These structures are similar to the distribution of $^{13}$CO(1-0) and $^{12}$CO(1-0) emissions of RCW38 \citep{Fukui2016}.
Figure~\ref{int-ci} shows the integrated intensity distribution (left and middle) and the declination-velocity diagram (right) of [CI] and $^{13}$CO(1-0) emission.
The [CI] cloud is composed of two velocity components:
Ring cloud ($v_{\rm LSR}$ = $-$8.3--9.1 km s$^{-1}$) and Finger cloud ($v_{\rm LSR}$ = 9.1--17.5 km s$^{-1}$),
which are generally similar to those of the CO clouds.
The distribution of $^{12}$CO(1-0) emission shows intermediate bridge structures between the Ring and Finger clouds (top panel of Figure 3 in \citet{Fukui2016}),
whereas the distribution of $^{13}$CO(1-0) and [CI] emissions does not trace such structures.
Therefore, the distribution of [CI] emission seems to be more similar to that of $^{13}$CO(1-0) emission than to that of $^{12}$CO(1-0) emission.
There are six peaks of [CI] and CO emissions in RCW38: four peaks in the Ring cloud and two peaks in the Finger cloud,
whereas the peak positions of [CI] emission are slightly different ($\sim$ 10$^{\prime \prime}$--30$^{\prime \prime}$) from those of $^{13}$CO(1-0) emission.
The velocity channel map of the [CI] emission of RCW38 is shown in Appendix \ref{sec:apd_1} (Figure \ref{channel_ci}).

\subsection{Spectrum of [CI] peaks}\label{sec:res_spec}
The top panels in Figure \ref{spectrum} show the spectra of [CI], $^{12}$CO(1-0), $^{13}$CO(1-0), and C$^{18}$O(1-0) emissions in the six [CI] peaks presented in Figure \ref{int-ci}.
In all the [CI] peaks of the Ring cloud, there are two or three peaks of $^{12}$CO(1-0) emission, which are caused by strong self-absorption \citep{Fukui2016}.
In all peaks, the overall velocity structure of [CI] emission is similar to that of $^{13}$CO(1-0) emission.
The [CI] and $^{13}$CO(1-0) emissions in the Ring cloud are also slightly affected by self-absorption.
In and around the Finger cloud, we did not detect C$^{18}$O(1-0) emission in any peak.
Table \ref{tbl_spectrum} shows the peak intensities and integrated intensities of the [CI] and CO emissions towards the peaks.

The bottom panels in Figure \ref{spectrum} show
the ratio of [CI]/$^{13}$CO(1-0) intensities ($I_{\rm [CI]}/I_{\rm ^{13}CO}$) of all peaks.
In the Ring cloud, the ratio around the velocity edge ($v_{\rm LSR}$ $\sim$ $-$4--$-$2 and/or 7--9 km s$^{-1}$) is higher than that around the velocity center,
except for the self-absorption velocity range (white circles in Figure \ref{spectrum}).
In the Finger cloud, the ratio around the large velocity ($v_{\rm LSR}$ $\sim$ 13--15 km s$^{-1}$) is higher than that around the small velocity ($v_{\rm LSR}\sim$ 11--13 km s$^{-1}$).

\subsection{Intensity ratio of [CI] to $^{13}$CO(1-0) emission}\label{sec:res_ratio}
Figure \ref{ratio-cico} shows the distribution of the ratio of [CI]/$^{13}$CO(1-0) integrated intensity (left and middle panels) and the declination-velocity diagram (right panel).
The ratio ($W_{\rm [CI]}/W_{\rm ^{13}CO}$) of the Ring and Finger clouds is 0.5--2.5 and 0.5--4.0, respectively.
As seen in the spectra (Figure~\ref{spectrum}), the ratio around the large velocity ($v_{\rm LSR}$ $\sim$ 13--15 km s$^{-1}$) of the Finger cloud is higher than that around the small velocity ($v_{\rm LSR}\sim$ 11--13 km s$^{-1}$).
For the Ring cloud, the ratio around the velocity edge ($v_{\rm LSR}$ $\sim$ $-$4--2 and/or 7--9 km s$^{-1}$) in the middle part of the declination
(decl. $\sim$ $-$47$^{\circ}$33$^{\prime}$00$^{\prime \prime}$--$-$47$^{\circ}$28$^{\prime}$00$^{\prime \prime}$)
is higher than that around the velocity center.
The ratio in the large and small parts of the declination (decl. $\geq$ $-$47$^{\circ}$28$^{\prime}$00$^{\prime \prime}$ and $\leq$ $-$47$^{\circ}$33$^{\prime}$00$^{\prime \prime}$)
of the Ring cloud is lower than that of the other region.
The velocity channel map of the intensity ratio of [CI]/$^{13}$CO(1-0) of RCW38 is shown in Appendix \ref{sec:apd_1} (Figure \ref{channel_cicoratio}).

\citet{Shimajiri2013} derived point-by-point correlations between the [CI] map and the other line maps
($^{12}$CO(1-0), $^{13}$CO(1-0), C$^{18}$O(1-0), and H$^{13}$CO$^{+}$) in the OMC-1 region in Orion-A cloud.
Consequently, they found that the [CI] emission has the highest correlation with the $^{13}$CO(1-0) emission.
Therefore, we also investigate the point-by-point correlations between the [CI] and $^{13}$CO (1-0) emissions
to compare the result of RCW38 with that of the OMC-1 region (Figure \ref{scatter-cico}).
We find that the Ring and Finger clouds also have  linear correlations between [CI] and $^{13}$CO(1-0) emissions, consistent with the result of the OMC-1 region.
From the least-squares fitting, the correlation of the Ring cloud, $I_{\rm ^{13}CO}$ = 0.64 ($\pm$ 0.00066) $I_{\rm [CI]}$,
is found to be very similar to that of the OMC-1 region presented in  \citet{Shimajiri2013}: $I_{\rm ^{13}CO}$ = 0.62 $I_{\rm [CI]}$.
On the other hand, the Finger cloud has a higher [CI]/$^{13}$CO(1-0) intensity ratio of $I_{\rm ^{13}CO}$ = 0.47 ($\pm$ 0.0043) $I_{\rm [CI]}$
than the Ring cloud and OMC-1 regions.

\subsection{Properties of RCW38}\label{sec:prop}
In this section, we derive the physical properties of the molecular clouds in RCW38 (excitation temperature, optical depth, and column density)
under the assumption of local thermodynamic equilibrium (LTE).

\subsubsection{Excitation temperature ($T_{\rm ex}$)}\label{sec:prop_tex}
We assumed that the excitation temperatures ($T_{\rm ex}$) of [CI] and $^{13}$CO(1-0) are identical to that of $^{12}$CO(1-0).
Since the critical densities of the [CI]($^3P_1$--$^3P_0$) and CO(1-0) lines are comparable ($n_{\rm cr}$ $\sim$ 10$^{3}$ cm$^{-3}$),
similar excitation conditions are expected for those lines (e.g. \cite{Ikeda2002}). 
This is also supported by the similar spatial and velocity structures of the [CI], $^{13}$CO, and $^{12}$CO emissions (see Section \ref{sec:res_dist}).
The $T_{\rm ex}$ of [CI] and CO transitions is derived from the peak intensity of the $^{12}$CO(1-0) emission
using the following equation:
\begin{equation}
T_{\rm ex} = \frac{h\nu /k}{\ln{\{1 + (h\nu /k) / (T(^{12}{\rm CO}) + J_\nu(T_{\rm BB}))\}}},
\end{equation}
where $T_{\rm BB}$ is the temperature of the cosmic background radiation (2.7 K) and $J_\nu$ is the radiation temperature, which is given by,
\begin{equation}
J_\nu (T) = \frac{h\nu /k}{{\rm exp}(h\nu /kT) - 1}.
\end{equation}
Using Eq. (2), Eq. (1) is written as follows:
\begin{equation}
T_{\rm ex} = \frac{5.53}{\ln{\{1 + 5.53 / (T(^{12}{\rm CO}) + 0.819)\}}} .
\end{equation}
Figure \ref{Tex} shows $T_{\rm ex}$ distribution for the Ring cloud (left panel) and the Finger cloud (right panel).
In the Ring cloud, $T_{\rm ex}$ around [CI] peaks rises up to $\sim$ 50 K.
In the Finger cloud, $T_{\rm ex}$ around [CI] peaks is lower ($\sim$ 25 K) than that in the Ring cloud.
However, $^{12}$CO(1-0) intensity of the Ring cloud is reduced by the strong self-absorption in the whole region of our [CI] mapping area (e.g., Figure \ref{spectrum}),
and therefore we might underestimate the $T_{\rm ex}$ of the whole region of the Ring cloud.
\citet{Fukui2016} investigated the kinetic temperature ($T_{\rm k}$) of RCW38 using large velocity gradient (LVG) analysis (e.g., \cite{Goldreich1974})
with $^{12}$CO(1-0), $^{12}$CO(3-2), and $^{13}$CO(1-0),
and found that the $T_{\rm k}$ in the outskirt of the Ring cloud is higher than 30--50 K (Figure 9 in \cite{Fukui2016}).
Therefore, we adopt a uniform $T_{\rm ex}$ of 49 K for the Ring cloud in the same manner as in \citet{Fukui2016}.
For the Finger cloud, we adopt a uniform $T_{\rm ex}$ of 25 K,
because the result of the LVG analysis from \citet{Fukui2016} shows that $T_{\rm k}$ of the peak and edge regions of the Finger cloud is higher than 20--30 K (Figure 10 in \citet{Fukui2016})
and the $^{12}$CO(1-0) emission at the edges of the Finger clouds is not optically thick because this region is more diffuse than the other regions of the cloud.
We examine the possible uncertainties caused by these assumptions in section \ref{sec:dis_abd}.
In addition, we present the results by employing a non-uniform distribution of $T_{\rm ex}$ (see Figure \ref{Tex}) in Appendix \ref{sec:apd_2}.

\subsubsection{Optical depth and column density}\label{sec:prop_den}
Under the assumption of LTE,
the total beam-averaged column density of [CI], $N({\rm [CI]})$, is given as follows:
\begin{eqnarray}
N({\rm [CI]}) &=& \frac{8\pi k \nu ^2}{3c^3hA_{10}} \times 
\int T_{\rm MB}\frac{\tau_{\rm [CI]}}{1 - e^{-\tau_{\rm [CI]}}}dv  \nonumber \\
&&\times
Q(T_{\rm ex}) {\rm exp}
\left(\frac{E_1}{kT_{\rm ex}}\right)
\left[1-\frac{J_\nu(T_{\rm BB})}{J_\nu(T_{\rm ex})}\right]^{-1},
\end{eqnarray}
where $Q(T_{\rm ex})$ is the ground-state partition function for the carbon atom,
$A_{10}$ is the Einstein $A$ coefficient for the $^3P_1$--$^3P_0$ transition, and $\tau$ is the optical depth.
The factor ($8\pi k \nu ^2$)/($3c^3hA_{10}$) is calculated to be 1.98 $\times$ 10$^{15}$ cm$^{-2}$ (K km s$^{-1}$)$^{-1}$.
The partition function is described as
\begin{equation}
Q(T_{\rm ex}) = 1 + 3 \exp\left(-\frac{E_1}{kT_{\rm ex}}\right) + 5 \exp\left(-\frac{E_2}{kT_{\rm ex}}\right),
\end{equation}
where $E_1$ and  $E_2$ represent the energies of the $J$ = 1 level (${E_1}/{k}$ = 23.6 K) and the $J$ = 2 level (${E_1}/{k}$ = 62.5 K).
The optical depth of [CI], $\tau_{\rm [CI]}$, is given by
\begin{equation}
\tau = -\ln \left\{1- \frac{T_{\rm MB}}{J_\nu(T_{\rm ex}) - J_\nu(T_{\rm BB})} \right\} .
\end{equation}
Using Eqs. (2), (5), and (6), Eq. (4) is written as follows:
\begin{equation}
N({\rm [CI]}) = 4.67 \times 10^{16} \frac{1 + 3e^{-23.6/T_{\rm ex}} + 5e^{-62.5/T_{\rm ex}}}{1 - e^{-23.6/T_{\rm ex}}} \int \tau_{\rm [CI]} dv .
\end{equation}
On the other hand, the column density of CO, $N({\rm CO})$, is estimated using the optical depth of $^{13}$CO(1-0), $\tau_{\rm ^{13}CO}$, and the integrated intensity of the $^{13}$CO(1-0) emission:
\begin{eqnarray}
N({\rm CO}) &=& [^{12}{\rm CO}]/[^{13}{\rm CO}] \times 4.57 \times 10^{13} \int T_{\rm MB} \frac{\tau_{\rm ^{13}CO}}{1 - e^{-\tau_{\rm ^{13}CO}}}
 dv \nonumber \\
&& \times \left(T_{\rm ex} + \frac{hB}{3k}\right) {\rm exp}\left(\frac{h\nu}{kT_{\rm ex}}\right)\left[1-\frac{J_\nu(T_{\rm BB})}{J_\nu(T_{\rm ex})}
\right]^{-1} \nonumber \\
&=& [^{12}{\rm CO}]/[^{13}{\rm CO}] \times 2.42 \nonumber \\
&& \times 10^{14} \frac{T_{\rm ex}+0.87}{1 - e^{-5.29/T_{\rm ex}}} \int \tau_{\rm ^{13}CO} dv ,
\end{eqnarray}
where $B$ is the rigid rotor rotation constant, the abundance ratio of $^{12}$CO/$^{13}$CO ([$^{12}$CO]/[$^{13}$CO]) is assumed to be 77 \citep{Wilson1994},
and $\tau_{\rm ^{13}CO}$ is estimated by Eq. (6).
Eqs. (1)---(7) were quoted from \citet{Oka2001} and \citet{Ikeda2002}. Eq. (8) was quoted from \citet{Mangum2015}.

Figure \ref{tau} shows the optical depth of the [CI] ($\tau_{\rm [CI]}$) and $^{13}$CO ($\tau_{\rm ^{13}CO}$) distributions of the six [CI] peaks.
Further, $\tau_{\rm [CI]}$ and $\tau_{\rm ^{13}CO}$ are 0.1--0.6 and 0.1--0.3, respectively, suggesting optically thin emission.
However, we note that self-absorption structures are present in the [CI] and $^{13}$CO spectra around the [CI] peaks (Figure \ref{spectrum})
therefore, [CI] and $^{13}$CO around strong peaks might be moderately optically thick.
Therefore, the [CI] and CO column densities in the Ring cloud are considered to be slightly underestimated.
We examined the possible underestimation caused by the self-absorption in section \ref{sec:dis_abd}.
Table \ref{tbl_prop} summarizes the properties (such as temperature, optical depth, and column density) of the six [CI] peaks in the Ring and Finger clouds.
We estimated the error of column densities
by varying the assumed $T_{\rm ex}$ of the Ring and Finger clouds from 30 K to 80 K and from 20 K to 30 K, respectively.
The details of the fluctuations are discussed in Section \ref{sec:dis_abd}.

Figure \ref{CIH2_column} shows the [CI] (Top) and H$_2$ (Bottom) column density distributions of RCW38.
The H$_2$ column density is estimated under the assumption of the abundance ratio of H$_2$/$^{12}$CO = 10$^4$ (e.g., \cite{Frerking1982}; \cite{Leung1984}).
As a result, we estimated
the [CI] column density of the Ring and Finger clouds to be (0.1--1.4) $\times$ 10$^{18}$ and (0.1--0.2) $\times$ 10$^{18}$ cm$^{-2}$, respectively,
and the H$_2$ column density of the Ring and Finger clouds to be (0.1--1.2) $\times$ 10$^{23}$ and $\sim$ 0.1 $\times$ 10$^{23}$ cm$^{-2}$, respectively.
Figure \ref{CICO_column} shows the ratio of [CI]/CO column density ($N_{\rm [CI]}$/$N_{\rm CO}$) distribution of RCW38.
The ratio in the Ring and Finger clouds are 0.05--0.15 and 0.1--0.5, respectively.

\subsubsection{[CI] - H$_2$ conversion factor}\label{sec:prop_con}
Figure \ref{CIint-H2column} shows the relationship between the [CI] integrated intensity ($W_{\rm [CI]}$) and H$_2$ column density ($N_{\rm H_2}$) derived from the $^{13}$CO(1-0) emission.
We also calculated $A_V$ using the relationship between the H$_2$ column density and the visual extinction: 
$N_{\rm H_2}$/$A_V$ = 9.4 $\times$ 10$^{20}$ cm$^{-2}$ mag$^{-1}$ \citep{Frerking1982} and put $A_V$ into Figure \ref{CIint-H2column}.
Both the Ring and Finger clouds have a linear relationship between $W_{\rm [CI]}$ and $N_{\rm H_2}$.
To estimate the conversion factor from $W_{\rm [CI]}$ to $N_{\rm H_2}$ ($X_{\rm [CI]}$  = $N_{\rm H_2}$/$W_{\rm [CI]}$), we used the least-squares fitting method for
the Ring cloud ($N_{\rm H_2}$ $\gtrsim$ 10$^{22}$ cm$^{-2}$; $A_V$  $\gtrsim$ 10 mag) and
the Finger cloud ($N_{\rm H_2}$ $\lesssim$ 10$^{22}$ cm$^{-2}$; $A_V$  $\lesssim$ 10 mag).
The conversion factors of the Ring and Finger clouds are
$X_{\rm [CI]}$ = 1.4 ($\pm$ 0.004) $\times$ 10$^{21}$ cm$^{-2}$ K$^{-1}$ km$^{-1}$ s (corresponding to $W_{\rm [CI]}$ = 7.4 ($\pm$ 0.02) $\times$ 10$^{-22}$ $N_{\rm H_2}$) and
$X_{\rm [CI]}$ = 6.3 ($\pm$ 0.08) $\times$ 10$^{20}$ cm$^{-2}$ K$^{-1}$ km$^{-1}$ s (corresponding to $W_{\rm [CI]}$ = 1.6 ($\pm$ 0.02) $\times$ 10$^{-21}$ $N_{\rm H_2}$),
respectively.

\citet{Offner2014} calculated $X_{\rm [CI]}$ for typical molecular clouds with low-UV ($G_0$ = 1 and 10) in the Milky Way using numerical simulations.
They demonstrated that [CI] is a good tracer of molecular gas distribution for column densities of up to 6 $\times$ 10$^{23}$ cm $^{-2}$ (corresponding to $A_V$ $\sim$ 600),
and derived the average $X_{\rm [CI]}$ of 1.1 $\times$ 10$^{21}$ cm$^{-2}$ K$^{-1}$ km$^{-1}$ s (corresponding to $W_{\rm [CI]}$ = 9.1 $\times$ 10$^{-22}$ $N_{\rm H_2}$).
They also reported that $X_{\rm [CI]}$ increases strongly with $N_{\rm H_2}$:
from $X_{\rm [CI]}$ $\sim$ 10$^{21}$ cm$^{-2}$ K$^{-1}$ km$^{-1}$ s at $N_{\rm H_2}$ = 10$^{21}$ cm$^{-2}$
to $X_{\rm [CI]}$ $\sim$ 10$^{22}$ cm$^{-2}$ K$^{-1}$ km$^{-1}$ s at $N_{\rm H_2}$ = 10$^{23}$ cm$^{-2}$ (see Figure 3 in \cite{Offner2014} for $G_0$ = 10),
which is consistent with our results (that $X_{\rm [CI]}$ of the Ring cloud is larger than that of the Finger cloud).
We note that the UV intensities used for the simulation are significantly smaller than those in the massive star-forming regions, such as the Orion and the Carina nebulae ($G_0$ $\sim$ 10$^4$--10$^5$)
similar to RCW38.
However, the intensity is considered to be attenuated inside molecular cloud, and
the intensities inside the RCW38 clouds are therefore similar to those used in the simulation.

\section{Discussion}\label{sec:dis}
\subsection{Relation between [CI] and $^{13}$CO(1-0) emission}\label{sec:dis_rel}
\begin{figure*}
 \begin{center}
  \includegraphics[width=16cm]{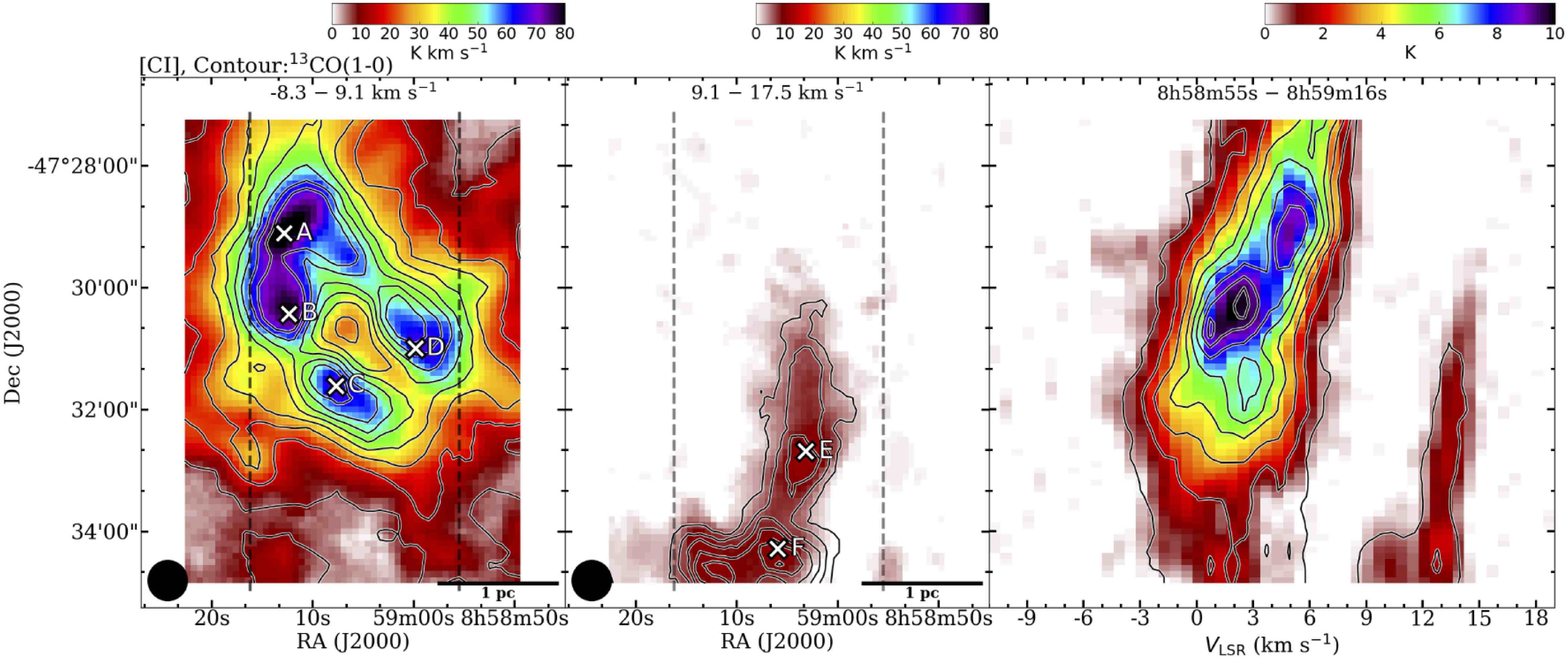} 
 \end{center}
\caption{[CI] integrated intensity distributions of RCW 38 (Left: Ring cloud, Middle: Finger cloud) and declination-velocity diagram (Right).
The black contours show the integrated intensity distribution and declination-velocity diagram of the $^{13}$CO(1-0) emission.
Contour levels of the Ring cloud are 3$\sigma$, 13$\sigma$, 23$\sigma$, 33$\sigma$, 43$\sigma$, 53$\sigma$, 63$\sigma$, 73$\sigma$, and 83$\sigma$ (1$\sigma$ = 0.5 K km s$^{-1}$).
Contour levels of the Finger cloud are 3$\sigma$, 5$\sigma$, 7$\sigma$, 9$\sigma$, 11$\sigma$, 13$\sigma$, 15$\sigma$, and 17$\sigma$ (1$\sigma$ = 0.4 K km s$^{-1}$).
Contour levels of the declination-velocity diagram are 3$\sigma$, 13$\sigma$, 23$\sigma$, 33$\sigma$, 43$\sigma$, 53$\sigma$, 63$\sigma$, and 73$\sigma$ (1$\sigma$ = 0.09 K).
The vertical dashed lines indicate the integration range in the declination-velocity diagram.
The ASTE [CI] data were smoothed to an HPBW of 40$^{\prime \prime}$ with a 2D Gaussian function for comparison with the Mopra $^{13}$CO(1-0) data.
The black filled circles at the lower left corners show the HPBW of the [CI] and the CO data (40$^{\prime \prime}$).
The six white cross symbols (designated from A to F) indicate the position of the [CI] peaks.
}
\label{int-ci}
\end{figure*}

\begin{figure*}
 \begin{center}
  \includegraphics[width=16cm]{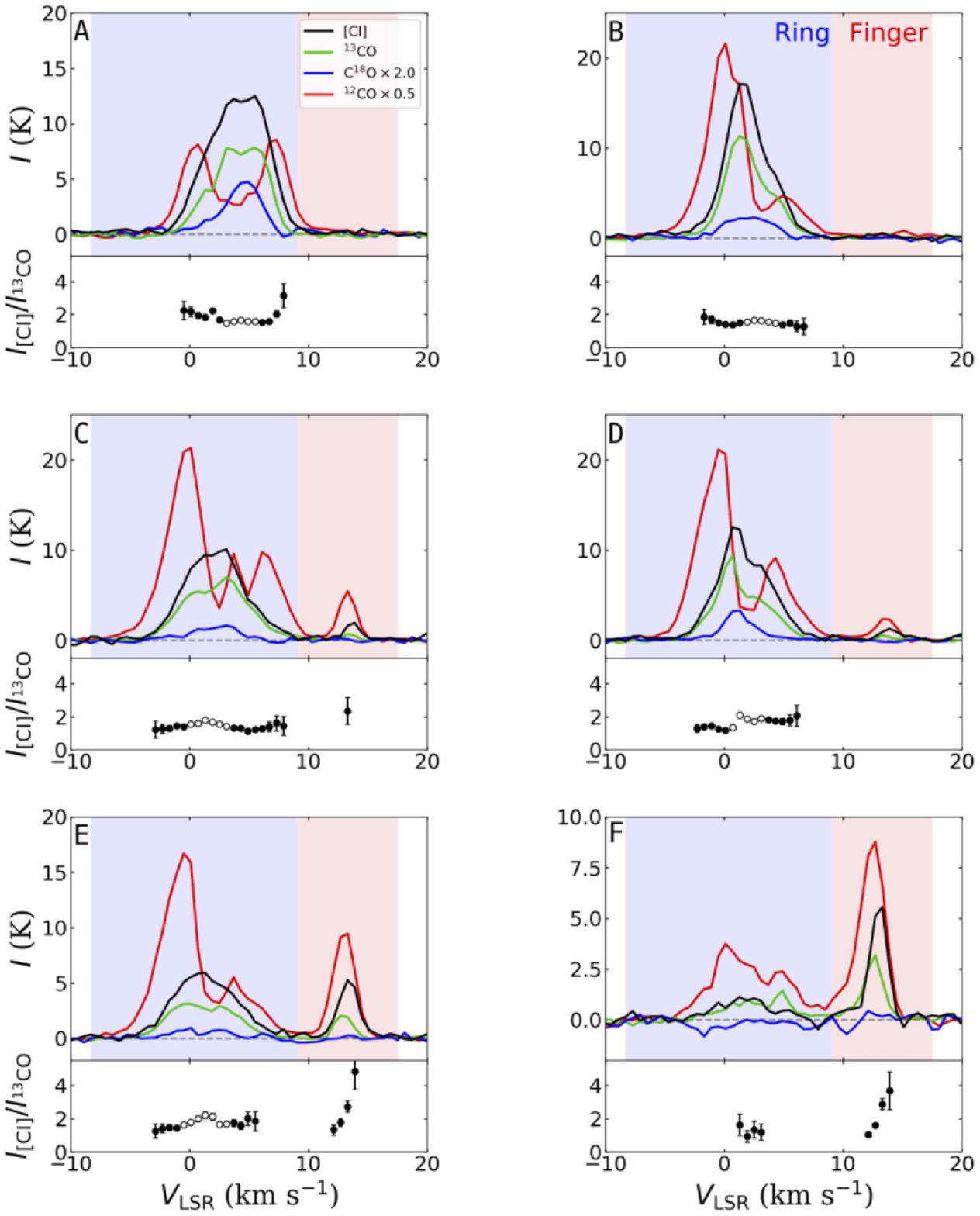} 
 \end{center}
\caption{
Averaged spectra of the [CI] and CO emissions for the six [CI] peaks detected. 
The spectra were obtained by averaging the intensities in a circle centered at the peak positions and with a radius of 20$^{\prime \prime}$,
corresponding to the HPBW of the [CI] and CO data.
[CI], $^{13}$CO(1-0), C$^{18}$O(1-0), and $^{12}$CO(1-0) are indicated by black, green, blue, and red lines, respectively.
Intensities of the C$^{18}$O(1-0) and $^{12}$CO(1-0) are multiplied by 2.0 and 0.5, respectively.
The blue and red areas indicate the velocity range of the Ring cloud (-8.3--9.1 km s$^{-1}$) and the Finger cloud (9.1--17.5 km s$^{-1}$), respectively.
The ratio of the [CI]/$^{13}$CO intensities ($I_{\rm [CI]}/I_{\rm ^{13}CO}$) for the six [CI] peaks is presented at the bottom of each spectrum.
These ratios are only derived in  $v_{\rm LSR}$ with $I_{\rm [CI]}$ $\geq$ 3$\sigma_{\rm [CI]}$ and $I_{\rm ^{13}CO}$ $\geq$ 3$\sigma_{\rm ^{13}CO}$
(1$\sigma_{\rm [CI]}$ = 0.2 K, 1$\sigma_{\rm ^{13}CO}$ = 0.1 K).
The black and white circles indicate the ratios that are affected and not affected by self-absorption, respectively.
Error bars represent 1$\sigma$ value of the ratio.
}
\label{spectrum}
\end{figure*}

\begin{figure*}
 \begin{center}
  \includegraphics[width=16cm]{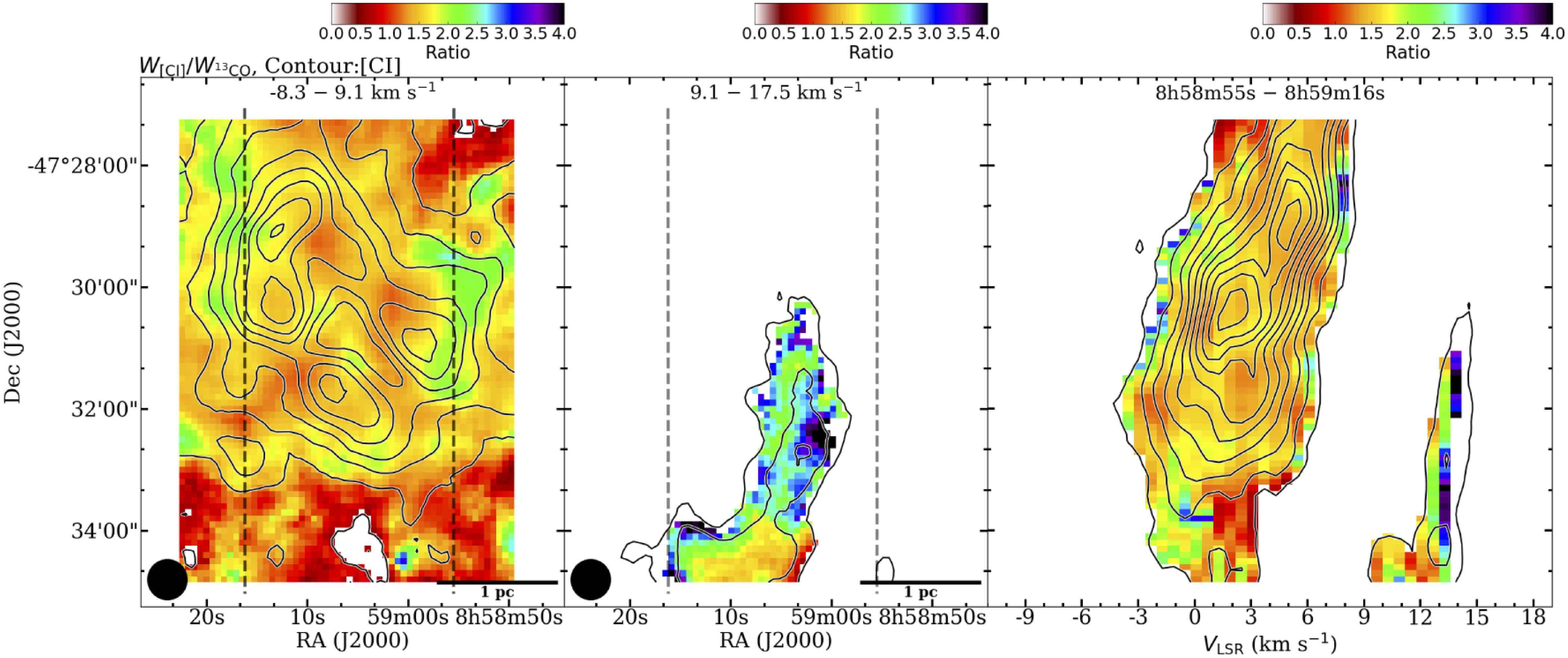} 
 \end{center}
\caption{Ratio of the [CI] / $^{13}$CO(1-0) integrated intensity ($W_{\rm [CI]}/W_{\rm ^{13}CO}$) distributions of RCW38 (Left: Ring cloud, Middle :Finger cloud) and declination-velocity diagram (Right).
The black contours show the integrated intensity distribution and declination-velocity diagram of the [CI] emission.
Contour levels of the Ring cloud are 3$\sigma$, 13$\sigma$, 23$\sigma$, 33$\sigma$, 43$\sigma$, 53$\sigma$, 63$\sigma$, 73$\sigma$, and 83$\sigma$ (1$\sigma$ = 1.0 km s$^{-1}$). 
Contour levels of the Finger cloud are 3$\sigma$, 8$\sigma$, and 13$\sigma$ (1$\sigma$ = 0.9 km s$^{-1}$).
Contour levels of the declination-velocity diagram are 3$\sigma$, 8$\sigma$, 13$\sigma$, 18$\sigma$, 23$\sigma$, 28$\sigma$, 33$\sigma$, and 38$\sigma$ (1$\sigma$ = 0.2 K).
The vertical dashed lines indicate the integration range in the declination-velocity diagram.
The black filled circles at the lower left corners show the HPBW of the [CI] and CO data (40$^{\prime \prime}$).
}
\label{ratio-cico}
\end{figure*}

\begin{figure*}
 \begin{center}
  \includegraphics[width=12cm]{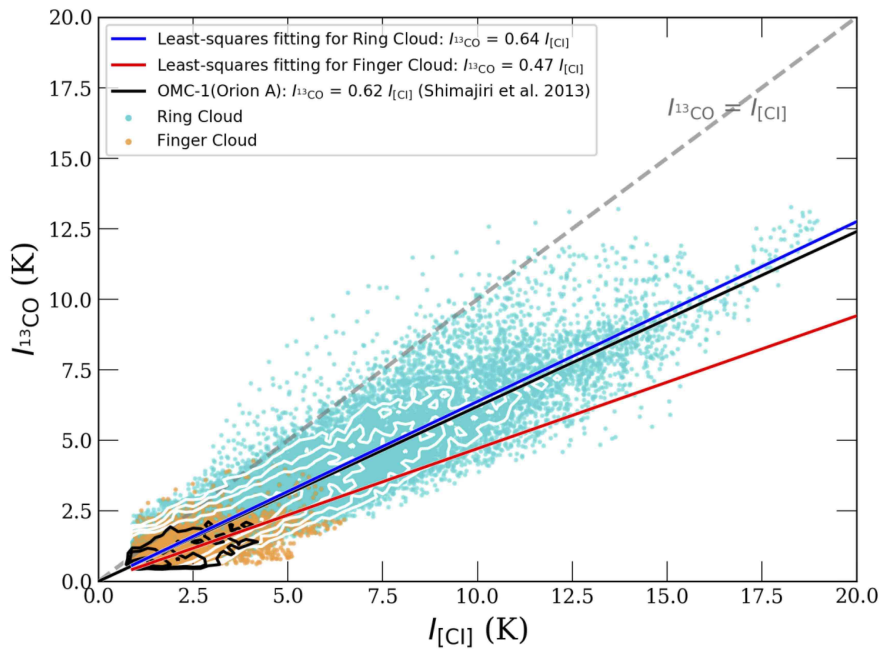} 
 \end{center}
\caption{Point-by-point correlations between [CI] and $^{13}$CO (Cyan: Ring cloud, Orange: Finger cloud).
Only pixels having intensities larger than the 3$\sigma$ noise levels in the data cube are plotted.
The 1$\sigma$ noise levels of the [CI] and $^{13}$CO are 0.3 and 0.2 K, respectively.
The blue (Ring cloud) and red (Finger cloud) lines show the results of leat-squares fitting:  $I_{\rm 13CO}$ = 0.64 ($\pm$ 0.00066) $\times$ $I_{\rm CI}$
and $I_{\rm 13CO}$ = 0.47 ($\pm$ 0.0043) $\times$ $I_{\rm CI}$, respectively.
The black line shows the result of leat-squares fitting in the OMC-1 region in Orion A cloud: $I_{\rm 13CO}$ = 0.62 $\times$ $I_{\rm CI}$ \citep{Shimajiri2013}.
The gray dotted-line represents $I_{\rm 13CO}$ =  $I_{\rm CI}$.
The white and black contours show the distributions of all pixels for the Ring cloud and the Finger cloud, respectively
(Ring cloud: 10, 20, 40, and 80 independent data points per 0.04 cell, Finger cloud: 10, 20, and 40 independent data points per 0.04 cell).
}
\label{scatter-cico}
\end{figure*}

\begin{figure*}
 \begin{center}
  \includegraphics[width=16cm]{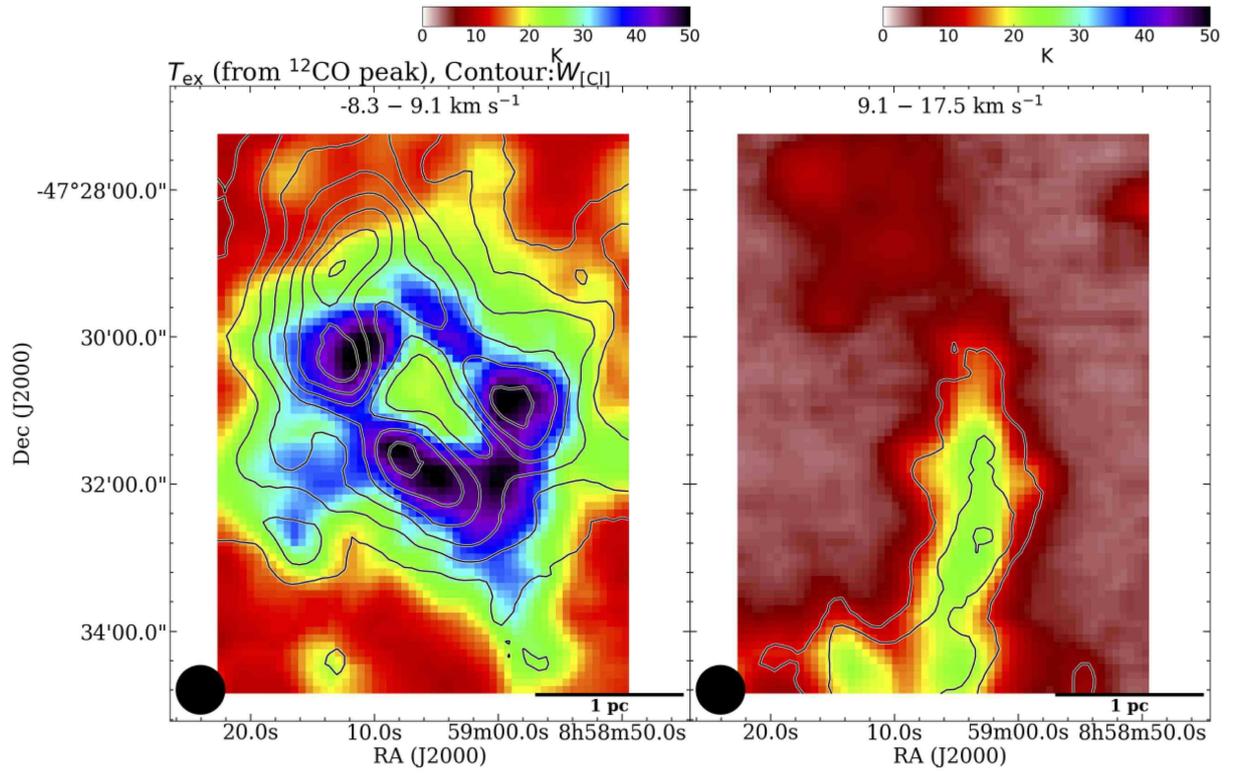}
 \end{center}
\caption{
Peak excitation temperature ($T_{\rm ex}$) distribution of RCW38 derived from the peak intensity of $^{12}$CO(1-0) data (Left: Ring cloud, Right: Finger cloud).
The black contours show the integrated intensity distribution and declination-velocity diagram of [CI] emission.
Contour levels of the Ring cloud are 3$\sigma$, 13$\sigma$, 23$\sigma$, 33$\sigma$, 43$\sigma$, 53$\sigma$, 63$\sigma$, 73$\sigma$, and 83$\sigma$ (1$\sigma$ = 1.0 km s$^{-1}$). 
Contour levels of the Finger cloud are 3$\sigma$, 8$\sigma$, and 13$\sigma$ (1$\sigma$ = 0.9 km s$^{-1}$).
The black filled circles at the lower left corners show the HPBW of the [CI] and CO data (40$^{\prime \prime}$).
}
\label{Tex}
\end{figure*}

\begin{figure*}
 \begin{center}
  \includegraphics[width=16cm]{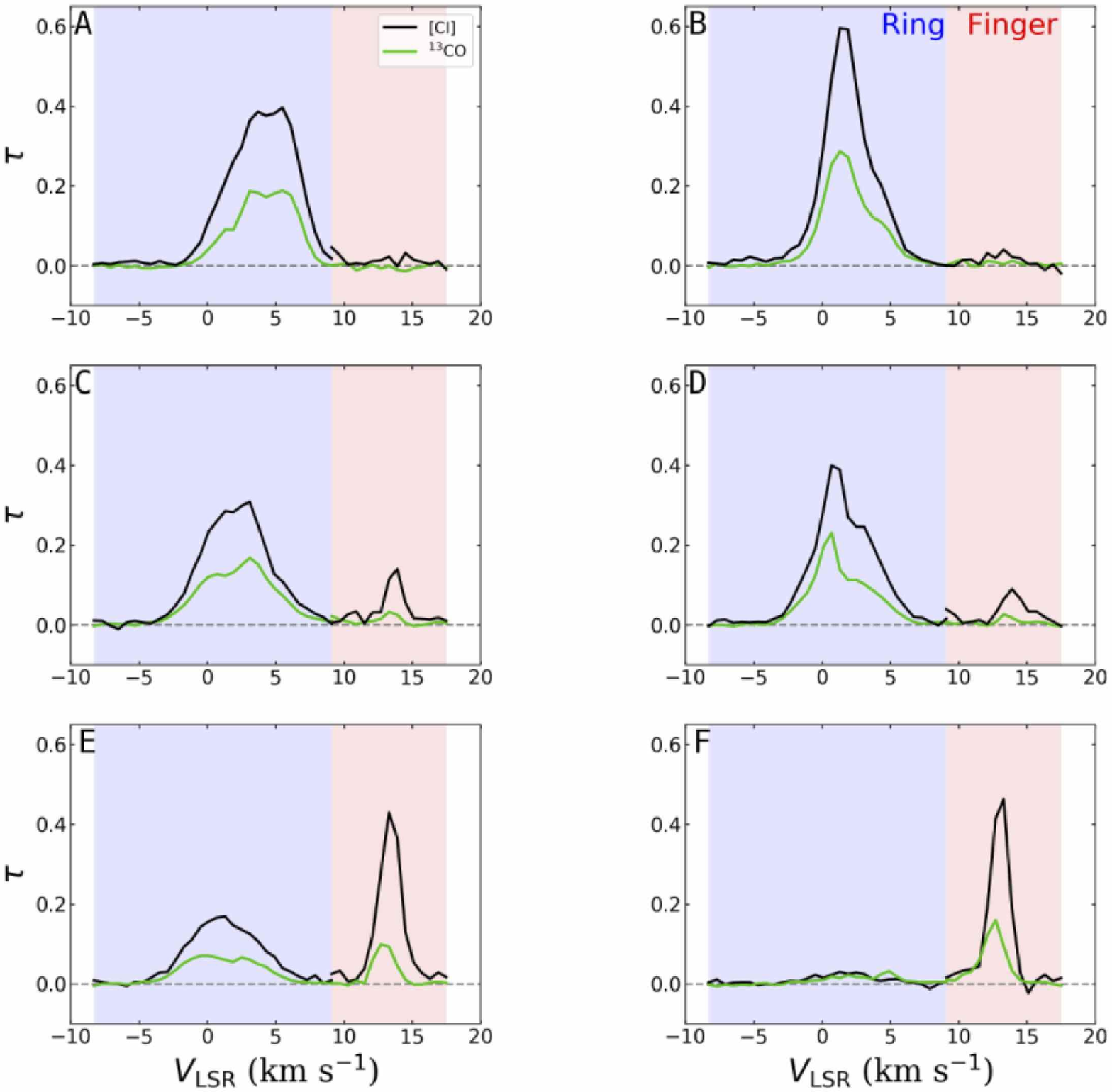} 
 \end{center}
\caption{
Optical depth of [CI] (black lines) and $^{13}$CO(1-0) (green lines) in the six [CI] peaks derived from the [CI] and $^{13}$CO(1-0) spectra in Figure \ref{spectrum}.
These optical depths are derived with Eq. (6) by assuming that the excitation temperature for the Ring and Finger clouds are 49 K and 25 K, respectively.
The blue and red areas indicate the velocity range of the Ring cloud (-8.3--9.1 km s$^{-1}$) and the Finger cloud (9.1--17.5 km s$^{-1}$), respectively.
}
\label{tau}
\end{figure*}

\begin{figure*}
 \begin{center}
  \includegraphics[width=16cm]{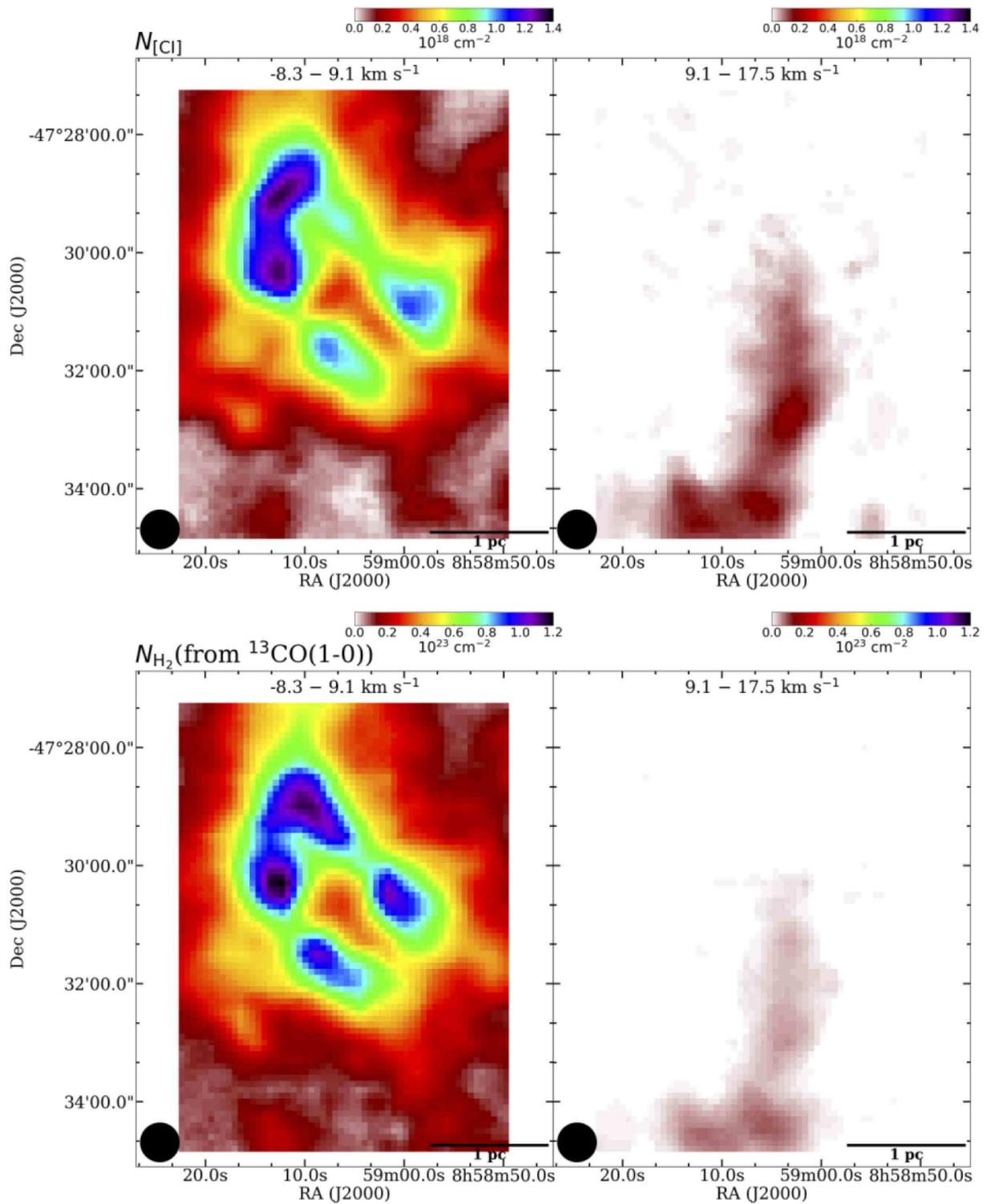} 
 \end{center}
\caption{[CI] (Top) and H$_2$ (Bottom) column density distributions of RCW38 (Left: Ring cloud, Right: Finger cloud).
The black filled circles at the lower left corners show the HPBW of the [CI] and CO data (40$^{\prime \prime}$).
}
\label{CIH2_column}
\end{figure*}

\begin{figure*}
 \begin{center}
  \includegraphics[width=16cm]{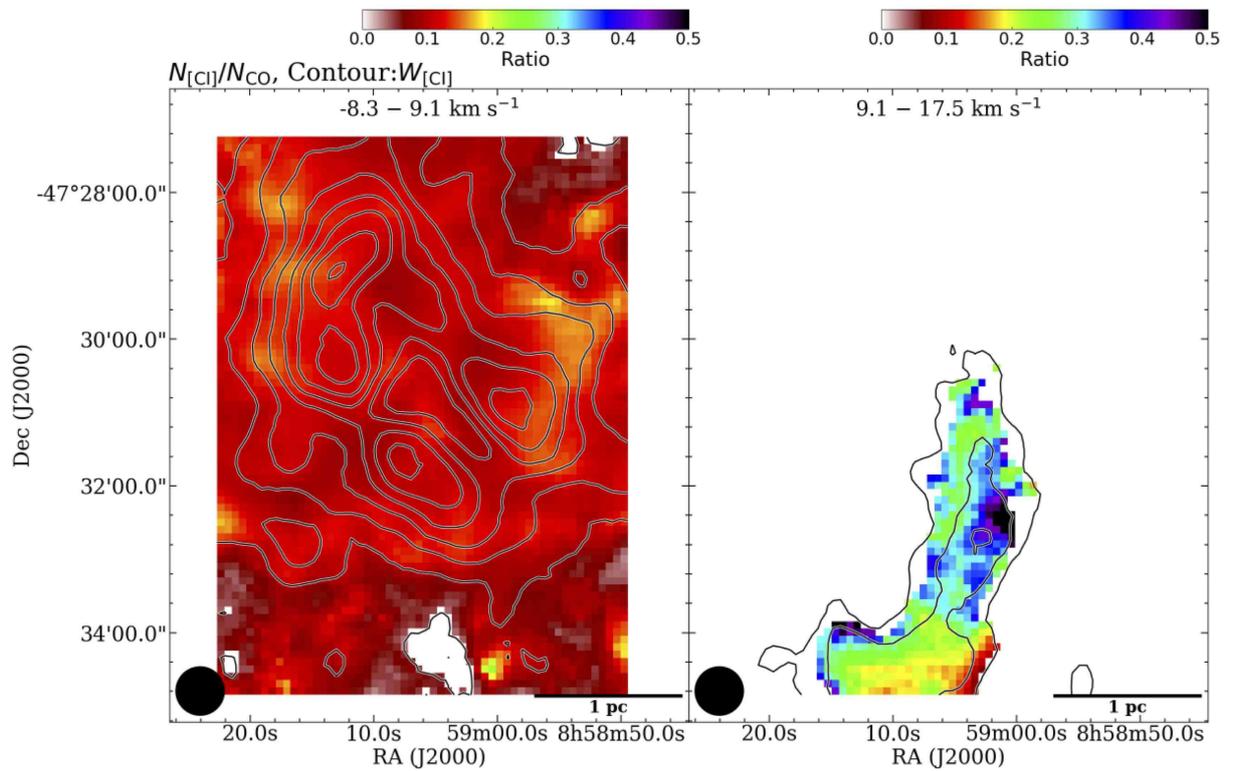}
 \end{center}
\caption{Ratio of [CI]/CO column density ($N_{\rm [CI]}/N_{\rm CO}$) distribution of RCW38 (Left: Ring cloud, Right: Finger cloud).
The black contours show the integrated intensity distribution of [CI] emission.
Contour levels of the Ring cloud are 3$\sigma$, 13$\sigma$, 23$\sigma$, 33$\sigma$, 43$\sigma$, 53$\sigma$, 63$\sigma$, 73$\sigma$, and 83$\sigma$ (1$\sigma$ = 1.0 km s$^{-1}$). 
Contour levels of the Finger cloud are 3$\sigma$, 8$\sigma$, and 13$\sigma$ (1$\sigma$ = 0.9 km s$^{-1}$).
The black filled circles at the lower left corners show the HPBW of the [CI] and CO data (40$^{\prime \prime}$).
}
\label{CICO_column}
\end{figure*}

\begin{figure*}
 \begin{center}
  \includegraphics[width=16cm]{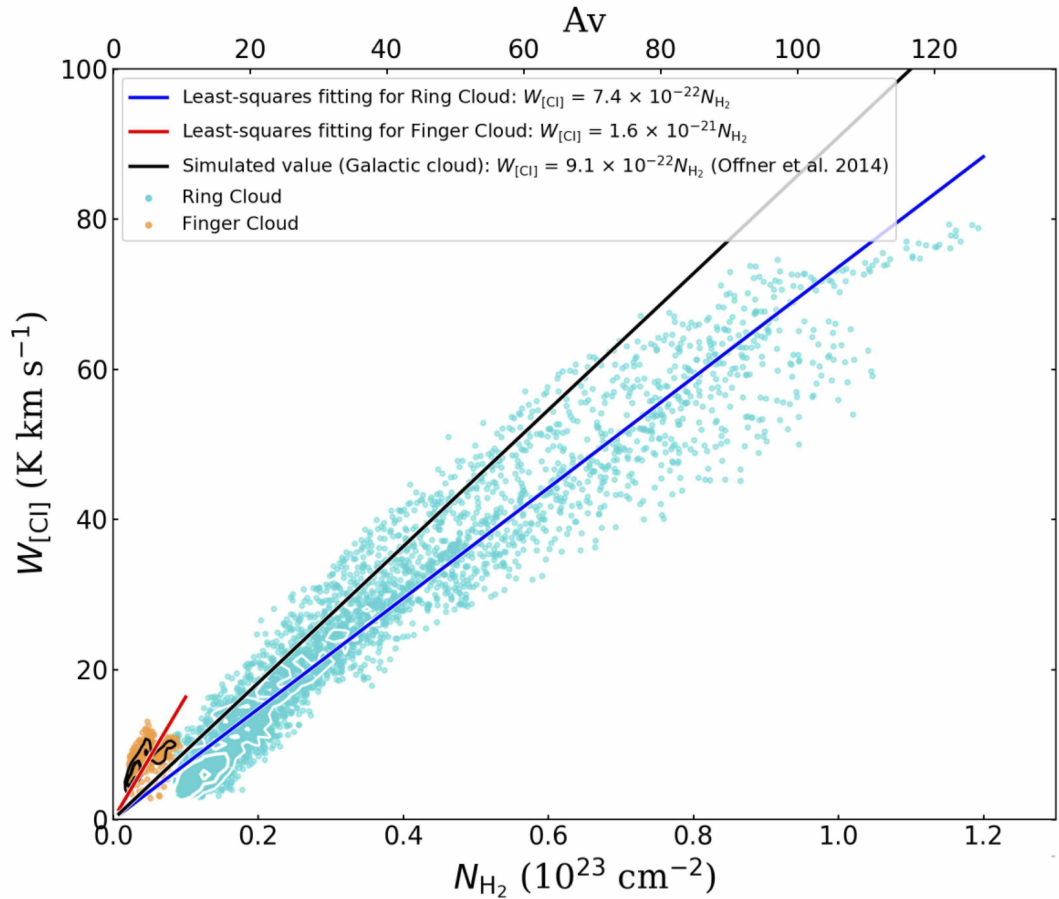} 
 \end{center}
\caption{
Point-by-point correlations between the integrated intensity of [CI] emission and the H$_2$ column density (Cyan: Ring cloud, Orange: Finger cloud).
Only pixels having intensities larger than the 3$\sigma$ noise levels in the data are plotted.
The 1 $\sigma$ noise levels of the [CI] integrated intensity and H$_2$ column density of the Ring cloud are 1.0 K km s$^{-1}$ and 1.2 $\times$ 10$^{21}$ cm$^{-2}$, respectively.
The 1 $\sigma$ noise levels of the [CI] integrated intensity and H$_2$ column density of the Finger cloud are 0.9 K km s$^{-1}$ and 5.6 $\times$ 10$^{20}$ cm$^{-2}$, respectively.
The blue and red lines show the results of least-squares fitting for the Ring cloud and the Finger cloud: $W_{\rm [CI]}$ = 7.4 ($\pm$ 0.02) $\times$ 10$^{-22}$ $N_{\rm{H_2}}$
and $W_{\rm [CI]}$ = 1.6 ($\pm$ 0.02) $\times$ 10$^{-21}$ $N_{\rm{H_2}}$, respectively.
The black line shows the simulated value for a typical Galactic cloud: $W_{\rm [CI]}$ = 9.1 $\times$ 10$^{-22}$ $N_{\rm{H_2}}$ by \citet{Offner2014}.
The white and black contours show the distributions of all pixels for the Ring cloud and the Finger cloud, respectively
(Ring cloud: 10 and 20 independent data points per 1.5 $\times$ 10$^{21}$ cell Finger cloud: 5 and 15 independent data points per 0.5 $\times$ 10$^{21}$ cell).
}
\label{CIint-H2column}
\end{figure*}

From our [CI] observations and previous CO observations \citep{Fukui2016} for RCW38,
we find that the spatial and velocity distributions of the [CI] emission are very similar to those of the $^{13}$CO(1-0) emission in RCW38 (Figures \ref{int-ci} and \ref{spectrum}).
These results are consistent with those from past [CI] observations (e.g., \cite{Ikeda2002,Shimajiri2013}).
In particular, the ratio of [CI]/$^{13}$CO(1-0) intensity ($I_{\rm [CI]}$/$I_{\rm ^{13}CO}$) of the Ring cloud is the same as that of the OMC-1 region in the Orion A cloud \citep{Shimajiri2013}.
These ratios are lower than that of the Finger cloud (Figure \ref{scatter-cico}).
The integrated intensity ratio ($W_{\rm [CI]}$/$W_{\rm ^{13}CO}$) of the Ring cloud is also lower than that of the Finger cloud (Figure \ref{ratio-cico}).
This trend may be related to the lower H$_2$ column density of the Finger cloud ($N_{\rm H_2}$ $\sim$ 10$^{21}$--10$^{22}$ cm$^{-2}$),
which is smaller than that of the Ring cloud and the Orion A cloud ($N_{\rm H_2}$ $\sim$ 10$^{22}$--10$^{23}$ cm$^{-2}$) by about an order of magnitude.
The Finger cloud may be affected more by UV radiation due to the small H$_2$ column density.
However, we note that a small H$_2$ column density does not always produce a high-ratio of  $W_{\rm [CI]}$/$W_{\rm ^{13}CO}$.
Figure \ref{ratio-cico} shows the low-ratio of $W_{\rm [CI]}$/$W_{\rm ^{13}CO}$ ($\sim$ 0.5--1.5) in the outer region of the Ring cloud, which has lower H$_2$ column density
($N_{\rm H_2}$ $\sim$ 10$^{22}$ cm$^{-2}$ in the bottom panel of Figure \ref{CIH2_column})
than in the inside region (ratio: $\sim$ 1.5--2.5) of the Ring cloud ($N_{\rm H_2}$ $\sim$ 10$^{23}$ cm$^{-2}$ in the bottom panel of Figure \ref{CIH2_column}).
It might be caused by the lower density of UV photons in the outer region because this region is far from the central O-star (IRS 2) and
is more strongly shielded than the dense molecular gas inside the Ring cloud.
This is consistent with a lower $^{12}$CO(3-2)/$^{12}$CO(1-0) intensity ratio ($\sim$ 0.3), suggesting a lower temperature in the outer region of the Ring cloud
than in the inside region (ratio of $\sim$ 0.5--1.5 in Figure \ref{Ring_region}; \cite{Fukui2016}).

\subsection{Relation between ratio of [CI]/[CO] abundance and H$_2$ column density}\label{sec:dis_abd}
The top panel of Figure \ref{CICO_H2column} shows the relationship between the ratios of [CI]/$^{13}$CO(1-0) integrated intensity ($W_{\rm [CI]}/W_{\rm ^{13}CO}$) and H$_2$ column density ($N_{\rm H_2}$),
and the bottom panel shows the relationship between the ratio of [CI]/CO column density ($N_{\rm [CI]}/N_{\rm CO}$) and $N_{\rm H_2}$.
We also show $A_V$, which is calculated from $N_{\rm H_2}$ \citep{Frerking1982} in Figure \ref{CICO_H2column}.
In the low-$A_V$ region ($A_V$ $\le$ 10 mag), $W_{\rm [CI]}/W_{\rm ^{13}CO}$ decreases with $A_V$ from $\sim$ 5 to $\sim$ 1.
In the high-$A_V$ region ($A_V$ $>$ 10 mag), $W_{\rm [CI]}/W_{\rm ^{13}CO}$ is nearly constant ($\sim$ 1.5) for $A_V$ of up to 100 mag.
A similar trend is seen in the relationship between $N_{\rm [CI]}/N_{\rm CO}$ and $A_V$ ($N_{\rm H_2}$).
In the low-$A_V$ region ($A_V$ $\le$ 10 mag), $N_{\rm [CI]}/N_{\rm CO}$ decreases with $A_V$ from $\sim$ 0.6 to $\sim$ 0.2.
In the high-$A_V$ region ($A_V$ $>$ 10 mag), $N_{\rm [CI]}/N_{\rm CO}$ is constant ($\sim$ 0.1) for $A_V$ of up to 100 mag.
Table \ref{tbl_std} summarizes the average and standard deviations of $W_{\rm [CI]}/W_{\rm ^{13}CO}$ and $N_{\rm [CI]}/N_{\rm CO}$ for the Ring and Finger clouds.
This shows that a large amount of [CI] with $\sim$ 10\% of CO abundance exists even in the high column density region of $A_V$ up to 100 mag.
This result is generally consistent with that of Orion A and B clouds by \citet{Ikeda2002},
but the method is slightly different from the present method.
We recalculated $N_{\rm [CI]}/N_{\rm CO}$ of Orion A and B clouds using a $^{12}$CO/$^{13}$CO abundance ratio of 77 \citep{Wilson1994},
instead of 60 adopted by \citet{Ikeda2002}.
In addition, we set the display range of $N_{\rm H_2}$ for Orion A and B clouds as 10$^{21}$ $\le$ $N_{\rm H_2}$ $\le$ 4.0 $\times$ 10$^{22}$ cm$^{-2}$,
which corresponds to a range of about 60\% of the whole data points in Figure 13 of \citet{Ikeda2002}.
The density of the Ring and Finger clouds were estimated to be larger than 10$^{3}$ cm$^{-3}$ according to the previous LVG analysis in Fukui et al. (2016).
In this case, the [$^{12}$CO]/[$^{13}$CO] abundance ratio may present spatial variations only over the relatively narrow clouds edge, which may be caused by selective photo-dissociation
(e.g. \cite{Warin1996,Szucs2014}).
Some chemical processes may also alter the [$^{12}$CO]/[$^{13}$CO] abundance ratio.
The simulation of \citet{Szucs2014} suggested that the chemical fractionation causes a factor of 2--3 decrease at intermediate cloud depth
(10$^{15}$ cm$^{-2}$ $\lesssim$ N($^{12}$CO) $\lesssim$ 10$^{17}$ cm$^{-2}$).
If we adopt a more realistic [$^{12}$CO]/[$^{13}$CO] distribution, the actual $N_{\rm [CI]}/N_{\rm CO}$ values can be slightly higher than our present derivations.
Thus, the uniform [$^{12}$CO]/[$^{13}$CO] is adopted in this paper provides the conservative lower limits of $N_{\rm [CI]}/N_{\rm CO}$.

\begin{figure*}
 \begin{center}
  \includegraphics[width=16cm]{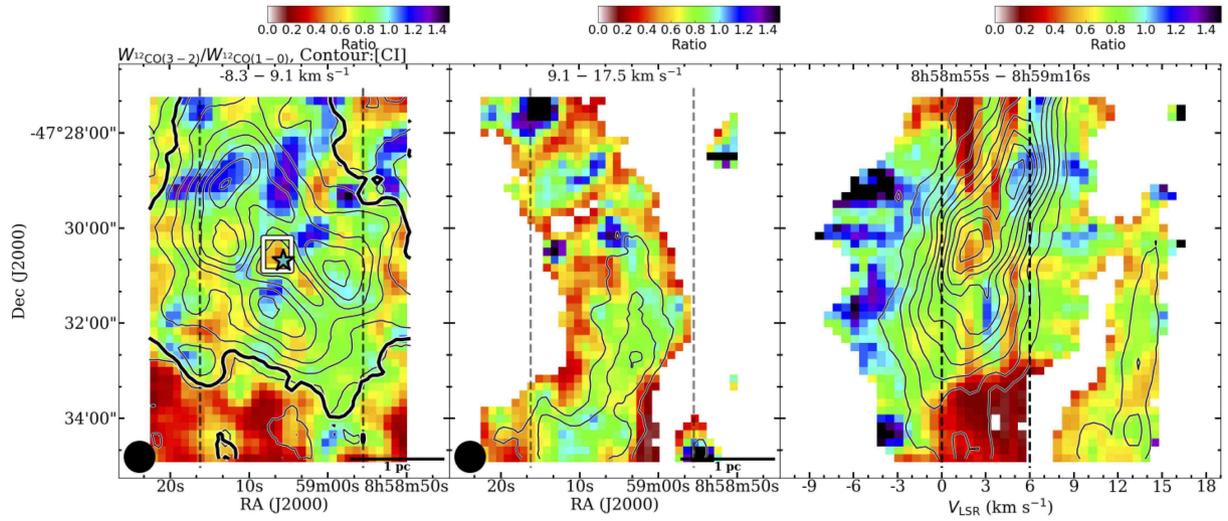} 
 \end{center}
\caption{
Ratio of $^{12}$CO(3-2)/$^{12}$CO(1-0) integrated intensity ($W_{\rm ^{12}CO(3-2)}$/$W_{\rm ^{12}CO(1-0)}$) distributions of RCW38 (Left: Ring cloud, Middle: Finger cloud) and declination-velocity diagram (Right). 
To remove the effect of the self-absorption, we except $v_{\rm LSR}$ = 0.0--6.0 km s$^{-1}$ component in the Left panel.
The thin black contours show the integrated intensity distribution and declination-velocity diagram of [CI] emission. 
Contour levels of the Ring cloud are 3$\sigma$, 13$\sigma$, 23$\sigma$, 33$\sigma$, 43$\sigma$, 53$\sigma$, 63$\sigma,$ 73$\sigma$, and 83$\sigma$ (1$\sigma$ = 1.0 km s$^{-1}$). 
Contour levels of the Finger cloud are 3$\sigma$, 8$\sigma$, and 13$\sigma$ (1$\sigma$ = 0.9 km s$^{-1}$). 
Contour levels of the declination-velocity diagram are 3$\sigma$, 8$\sigma$, 13$\sigma$, 18$\sigma$, 23$\sigma$, 28$\sigma$, 33$\sigma$, and 38$\sigma$ (1$\sigma$ = 0.2 K). 
The thick black contour in the left panel shows the 20$\sigma$ value (2.0 km s$^{-1}$) of the $^{13}$CO(1-0) integrated intensity.
The white box in the left panel shows the inner region of the Ring cloud.
The cyan star symbol indicates the peak of IRS 2.
The vertical dashed lines in the Left and Middle panels indicate the integration range in the declination-velocity diagram. 
The black vertical dashed line in the Right panel indicates the region strongly affected by the self-absorption.
The black filled circles at the lower left corners show the HPBW of the [CI] and CO data (40$^{\prime \prime}$).
}
\label{Ring_region}
\end{figure*}

Figure \ref{Ring_in_out} shows the relationship between $N_{\rm [CI]}/N_{\rm CO}$ and $N_{\rm H_2}$ for the inner, outer, and the other regions of the Ring cloud.
We defined the outer region of the Ring cloud by a low-integrated intensity of $^{13}$CO(1-0): lower than 10 K km s$^{-1}$, which corresponds to a value of 20$\sigma$ (black thick contour in Figure \ref{Ring_region}).
We defined the inner region of the Ring cloud as the central region around IRS 2 (white box in Figure \ref{Ring_region}).
The average ratio of $N_{\rm [CI]}/N_{\rm CO}$ (and standard deviation) in the outer, inner, and other regions is
0.090 (std: 0.086), 0.11 (std: 0.0083), and 0.11(std: 0.019), respectively.
The $N_{\rm [CI]}/N_{\rm CO}$ ratio in the outer region is lower than that in the other and inner regions except for the low-column density ($N_{\rm H_2}$ $\lesssim$ 10$^{22}$ cm$^{-2}$),
where the ratio becomes higher due to the diffuse distribution of molecular gas.
The $N_{\rm [CI]}/N_{\rm CO}$ ratio in the inner region is slightly higher than that in the other region, especially for the high-column density ($N_{\rm H_2}$ $\gtrsim$ 3 $\times$ 10$^{22}$ cm$^{-2}$).
The variation in the ratios suggest that the outer region is affected less and the inner region is affected more by UV photons than in the other region.
This result is consistent with the physical association between the Ring cloud and the cluster in RCW38 reported by \citet{Fukui2016}:
cluster is located in and ionizes the central cavity of the Ring cloud.

Next, we discuss the effect of uncertainty in $T_{\rm ex}$ on the $N_{\rm [CI]}/N_{\rm CO}$.
To assess the biases caused by self-absorption,
we re-constructed spectra corrected from the absorption
of $^{12}$CO(1-0), $^{13}$CO(1-0), and [CI] by fitting gaussian profile to them for 6 [CI] peaks (Figure \ref{Gauss_fit}).
The highest reconstructed $^{12}$CO(1-0) intensity corresponds to a $T_{\rm ex}$ of 80 K.
On the other hand, the previous LVG analysis derived that the lowest $T_{\rm ex}$ in the Ring cloud is $\sim$ 30 K \citep{Fukui2016}.
The uncertainties of $N_{\rm [CI]}/N_{\rm CO}$ in the Ring cloud can be gauged by changing the assumed $T_{\rm ex}$ from 30 K to 80 K.
Similarly,  to be consistent with the previous LVG analysis \citep{Fukui2016}, we changed the assumed $T_{\rm ex}$ from 20 K to 30 K to assess the uncertainty of $N_{\rm [CI]}/N_{\rm CO}$ in the Finger cloud.
From Eqs. (6)--(8) in Section \ref{sec:prop_den},
$N_{\rm [CI]}/N_{\rm CO}$ is written as follows:
\begin{equation}
N_{\rm [CI]}/N_{\rm CO} = \alpha R(\tau) f(T_{\rm ex}) ,
\end{equation}
\begin{equation}
\alpha = 2.51 ,
\end{equation}
\begin{eqnarray}
R(\tau) &=& \frac{\int \tau_{\rm [CI]} dv}{\int \tau_{\rm ^{13}CO} dv} \nonumber \\
 &=&
\frac{\int \ln \left\{1- \frac{T_{\rm MB}(\rm [CI])}{J_\nu(T_{\rm ex}) - J_\nu(T_{\rm BB})} \right\}dv}
{\int \ln \left\{1- \frac{T_{\rm MB}(\rm ^{13}CO)}{J_\nu(T_{\rm ex}) - J_\nu(T_{\rm BB})} \right\}dv} \nonumber \\
&=& R(T_{\rm MB}, T_{\rm ex}) ,
\end{eqnarray}
\begin{equation}
f(T_{\rm ex}) = \frac{(1 + 3e^{-23.6/T_{\rm ex}} + 5e^{-62.5/T_{\rm ex}})(1 - e^{-5.29/T_{\rm ex}})}{(1 - e^{-23.6/T_{\rm ex}})(T_{\rm ex}+0.87)} .
\end{equation}

For the Ring cloud, $R(T_{\rm MB}, T_{\rm ex})$ and $f(T_{\rm ex})$ change from $\sim$ 140\% to $\sim$ 90\% of $R(T_{\rm MB}, 49 {\rm K})$ and $\sim$ 130\% to $\sim$ 70\% of $f(49{\rm K})$, respectively,
with $T_{\rm ex}$ changing from 30 K to 80 K.
For the Finger cloud, $R(T_{\rm MB}, T_{\rm ex})$ and $f(T_{\rm ex})$ change from $\sim$ 120\% to $\sim$ 90\% of $R(T_{\rm MB}, 25 {\rm K})$ and $\sim$ 110\% to $\sim$ 90\% of $f(25K)$, respectively,
with $T_{\rm ex}$ changing from 20 K to 30 K.
Therefore, the possible value of $N_{\rm [CI]}/N_{\rm CO}$ in the Ring cloud is $\sim$ 0.2--0.06 by considering the uncertainty of $T_{\rm ex}$ (see error values in Table \ref{tbl_prop}).
The possible value of $N_{\rm [CI]}/N_{\rm CO}$ in the Finger cloud is higher than that in the Ring cloud ($\gtrsim$ 0.2).
 
Self-absorption may cause $N_{\rm [CI]}$ and $N_{\rm CO}$ to be underestimated.
For 5 [CI] peaks\footnote{
We could not detect any self-absorption features of [CI] and $^{13}$CO emission in one peak (labeled as F) out of the six [CI] peaks}, 
the $N_{\rm [CI]}$ and $N_{\rm CO}$ derived from the directly observed spectra are about 80--95 \% and 80--90 \% of those derived based on the gaussian-reconstructed [CI] and $^{13}$CO(1-0) spectra.
These results indicated that the self-absorption of [CI] and $^{13}$CO(1-0) have minor effect on $N_{\rm [CI]}$, $N_{\rm CO}$, and $N_{\rm [CI]}/N_{\rm CO}$.

Our main conclusions from this section are: 1) $N_{\rm [CI]}/N_{\rm CO}$ is approximately 0.1 for $A_V$ of up to 100 mag
and 2) variation in UV irradiation between each region of the Ring cloud (inner, outer, and the other regions) is detected,
and it is not dependent on our $T_{\rm ex}$
assumption (uniform $T_{\rm ex}$ for the Ring and Finger clouds).
We present the results in which $T_{\rm ex}$ is derived from the peak intensity of $^{12}$CO(1-0) (see Figure \ref{Tex}) in Appendix \ref{sec:apd_2}.

\subsection{PDR model}\label{sec:dis_pdr}
The present analysis shows that the [CI]/CO abundance ratio is $\sim$ 0.1 over a wide range of $A_V$ (10--100 mag),
whereas is varies from $\sim$ 0.5 to $\sim$ 0.1 for $A_V$ of less than 10 mag.
This behavior is generally consistent with that of the PDR model predicted by the classical layered distribution, whereas the large [CI] abundance at large $A_V$ is not consistent with the model (e.g. \cite{Rollig2007}).
The present results further evident the large residual [CI] in the deep interiors of a molecular cloud having $A_V$ of a few 10 to 100 mag,
which are consistent with the previous observations of the Orion and the other regions (e.g., \cite{Ikeda2002,Shimajiri2013,Burton2015}).
\citet{Glover2015} presented the model calculations of [CI] and CO in a highly turbulent cloud of $A_V$ $\sim$ 10 mag,
and showed that an inhomogeneous cloud allows the UV radiation to penetrate the cloud deeply at $A_V$ $\sim$ 10 mag.
This cloud is similar to a low column density cloud, such as Taurus (e.g., \cite{Mizuno1995}), whereas RCW38 is a giant molecular cloud (GMC) whose column density is 10 times larger than that of in Taurus.
This indicates that carbon ionizing photons of 11--13.6 eV deeply penetrate a GMC.
We suggest that such high [CI] abundance generally indicates a highly clumpy density distribution in a GMC.
The present resolution of the Mopra and ASTE telescopes, i.e., $\sim$ 0.3 pc, may be insufficient to resolve such clumps.
For a better understanding, we plan high-resolution [CI] observations toward RCW38 with ALMA.
The expected clumped distribution is in fact directly shown by the recent high-resolution C$^{18}$O image of RCW38 at 0.02 pc resolution obtained with ALMA \citep{Torii2019},
where dense cores of size 0.02--0.08 pc exhibit clumpy and filamentary distribution.
This distribution is compared with the magnetohydrodynamical numerical simulations of colliding molecular flows \citep{Inoue2013}.
By synthetic observations of the numerical simulations, \citet{Fukui2019} demonstrated that the core mass function and the size distribution indicate that the cores are sufficiently massive to form high mass stars,
suggesting that clumped distribution may be  
an important stage prior to a high mass star formation in a GMC.
It remains to be investigated further if the clumpy distribution of the [CI] gas in RCW38 was created because of cloud-cloud collision,
while the collision is a process which creates the clumpy distribution. 
We cautiously include an alternative possibility that the distribution may be a general property of the dense gas in a GMC, independent of the past cloud-collision episode.
To clarify this issue, we need to extend [CI] observations to other GMCs having different environments and histories.

\section{Summary}\label{sec:sum}
We present the results of [CI] observations toward the super star cluster RCW38 with the ASTE 10 m sub-mm telescope.
Our main results are as follows.
\begin{enumerate}
\item The overall distribution of [CI] emission is similar to that of $^{13}$CO(1-0) emission presented in \citet{Fukui2016},
including two velocity components:
Ring cloud ($v_{\rm LSR}$ = $-$8.3--9.1 km s$^{-1}$) and Finger cloud ($v_{\rm LSR}$ = 9.1--17.5 km s$^{-1}$).
These results are consistent with those of the past [CI] observations (e.g., \cite{Ikeda2002,Shimajiri2013}).

\item By assuming that the $T_{\rm ex}$ for the Ring and Finger clouds are 49 K and 25 K, respectively,
as derived from the $^{12}$CO(1-0) emission, the optical depth of the [CI] emission is found to be $\tau$ of 0.1--0.6,
suggesting mostly optically thin emission.
The column densities of [CI] of the Ring and Finger clouds are (0.1--1.4) $\times$ 10$^{18}$ and (0.1--0.2) $\times$ 10$^{18}$ cm$^{-2}$,
respectively.

\item An empirical conversion factor from the [CI] integrated intensity to the H$_2$ column density was estimated to be
$X_{\rm [CI]}$ = 6.3 $\times$ 10$^{20}$ cm$^{-2}$ K$^{-1}$ km$^{-1}$ s at $A_V$ of less than 10 mag, and
$X_{\rm [CI]}$ = 1.4 $\times$ 10$^{21}$ cm$^{-2}$ K$^{-1}$ km$^{-1}$ s for $A_V$ of 10--100 mag.

\item  The column density ratio of [CI] to CO ($N_{\rm [CI]}$/$N_{\rm CO}$) was estimated as $\sim$ 0.1 for $A_V$ of 10--100 mag.
This value is consistent with that of the Orion A and B clouds \citep{Ikeda2002}.
The present results encompass a large $A_V$ (of up to 100 mag) than that for Orion (which covers $A_V$ of up to $\sim$ 60 mag).

\item  The results are consistent with the highly clumped density distributions revealed by the recent ALMA observations of C$^{18}$O by \citet{Torii2019}.
Numerical simulations of colliding molecular gas flows show that dense cores of similar sizes are formed via compression,
which further corroborates the clumped distribution suggested by the high [CI] abundance in the deep interiors of GMC with $A_V$ of $\sim$100 mag.

\end{enumerate}
%%%%%FIGURE_and_TABLE%%%%
\clearpage

%==================table==================%
\begin{table*}
  \tbl{[CI] and CO peak intensities and integrated intensities of 6 [CI] peaks}{%
  \begin{tabular}{cccccccccc}
      \hline  \hline
      Name & R.A.(J2000) & Decl.(J2000) & Line & \multicolumn{2}{c}{$T_{\rm peak}$ (K)} & \multicolumn{2}{c}{$\int T_{\rm MB} dv$ (K km s$^{-1}$)} \\
      &  &  & & Ring & Finger & Ring & Finger \\ 
      \hline
     A & 8:59:12.86 & -47:29:06.52 & [CI] & 12.5 $\pm$ 0.3 & ---\footnotemark[$*$] & 78.0 $\pm$ 3.0 & --- \\
   	&                 &                    & $^{13}$CO(1-0) & 7.8 $\pm$ 0.2 & --- & 44.0 $\pm$ 2.0 & --- \\
   	&                 &                    & C$^{18}$O(1-0) & 2.4 $\pm$ 0.1 & --- & 9.1 $\pm$ 1.0 & --- \\
   	&                 &                    & $^{12}$CO(1-0) & 17.1 $\pm$ 0.4 & 4.1 $\pm$ 0.4 & 110.0 $\pm$ 5.0 & 3.7 $\pm$ 0.5 \\ \hline      
      B & 8:59:12.35 & -47:30:25.55 & [CI] & 17.1 $\pm$ 0.3 & --- & 73.0 $\pm$ 3.0 & --- \\
   	&                 &                    & $^{13}$CO(1-0) & 11.3 $\pm$ 0.2 & --- & 47.0 $\pm$ 2.0 & --- \\
   	&                 &                    & C$^{18}$O(1-0) & 1.2 $\pm$ 0.1 & --- & 5.4 $\pm$ 0.6 & --- \\
   	&                 &                    & $^{12}$CO(1-0) & 43.2 $\pm$ 0.4 & 1.7 $\pm$ 0.4 & 190.0 $\pm$ 6.0 & 1.8 $\pm$ 0.5 \\ \hline
      C & 8:59:07.70 & -47:31:36.79 & [CI] & 10.1 $\pm$ 0.3 & 2.0 $\pm$ 0.3 & 62.0 $\pm$ 3.0 & 2.1 $\pm$ 0.4 \\
   	&                 &                    & $^{13}$CO(1-0) & 7.0 $\pm$ 0.2 & 0.7 $\pm$ 0.2 & 43.0 $\pm$ 2.0 & 0.4 $\pm$ 0.1 \\
   	&                 &                    & C$^{18}$O(1-0) & 0.8 $\pm$ 0.1 & --- & 3.0 $\pm$ 0.5 & --- \\
   	&                 &                    & $^{12}$CO(1-0) & 42.7 $\pm$ 0.4 & 10.9 $\pm$ 0.4 & 240.0 $\pm$ 6.0 & 22.0 $\pm$ 2.0 \\ \hline
      D & 8:58:59.84 & -47:30:59.76 & [CI] & 12.6 $\pm$ 0.3 & 1.3 $\pm$ 0.3 & 59.0 $\pm$ 3.0 & 1.9 $\pm$ 0.5\\
   	&                 &                    & $^{13}$CO(1-0) & 9.4 $\pm$ 0.2 & --- & 38.0 $\pm$ 2.0 & --- \\
   	&                 &                    & C$^{18}$O(1-0) & 1.7 $\pm$ 0.1 & --- & 4.4 $\pm$ 0.4 & --- \\
   	&                 &                    & $^{12}$CO(1-0) & 42.3 $\pm$ 0.4 & 4.8 $\pm$ 0.4 & 200.0 $\pm$ 6.0 & 9.9 $\pm$ 1.0 \\ \hline
      E & 8:59:03.18 & -47:32:41.20 & [CI] & 6.0 $\pm$ 0.3 & 5.2 $\pm$ 0.3 & 37.0 $\pm$ 3.0 & 10.0 $\pm$ 0.9 \\
   	&                 &                    & $^{13}$CO(1-0) & 3.1 $\pm$ 0.2 & 2.0 $\pm$ 0.2 & 20.0 $\pm$ 2.0 & 3.7 $\pm$ 0.5 \\
   	&                 &                    & C$^{18}$O(1-0) & 0.5 $\pm$ 0.1 & --- & 1.3 $\pm$ 0.5 & --- \\
   	&                 &                    & $^{12}$CO(1-0) & 33.4 $\pm$ 0.4 & 18.9 $\pm$ 0.4 & 160.0 $\pm$ 6.0 & 45.0 $\pm$ 2.0 \\ \hline
      F & 8:59:05.97 & -47:34:17.29 & [CI] & 1.1 $\pm$ 0.3 & 5.6 $\pm$ 0.3 & 2.5 $\pm$ 0.7 & 9.5 $\pm$ 0.7 \\
   	&                 &                    & $^{13}$CO(1-0) & 1.4 $\pm$ 0.2 & 3.2 $\pm$ 0.2 & 4.5 $\pm$ 1.0 & 6.1 $\pm$ 0.7 \\
   	&                 &                    & C$^{18}$O(1-0) & --- & --- & --- & --- \\
   	&                 &                    & $^{12}$CO(1-0) & 7.5 $\pm$ 0.4 & 17.6 $\pm$ 0.4 & 46.0 $\pm$ 5.0 & 49.0 $\pm$ 2.0 \\
      \hline
    \end{tabular}}
\begin{tabnote}
      \footnotemark[$*$] $T_{\rm peak}$ is less than 3$\sigma$ (1$\sigma_{\rm [CI]}$ = 0.3 K, 1$\sigma_{\rm ^{13}CO(1-0)}$ = 0.2 K, 1$\sigma_{\rm C^{18}O(1-0)}$ = 0.1 K, 1$\sigma_{\rm ^{12}CO(1-0)}$ = 0.4 K)
\end{tabnote}
\label{tbl_spectrum}
\end{table*}
%==================table==================%
\clearpage

\clearpage

%==================table==================%
\begin{table*}
\tbl{Properties of 6 [CI] peaks}{%
\begin{tabular}{ccccccccc}
\hline  \hline
Cloud &Name & $T_{\rm ex}$ (K) & $\overline{\tau_{[CI]}}$\footnotemark[$*$] & $\overline{\tau_{\rm ^{13}CO}}$& $N_{\rm [CI]}$ (cm$^{-2}$)\footnotemark[$\dagger$] & $N_{\rm CO}$ (cm$^{-2}$) & $W_{\rm [CI]}/W_{\rm ^{13}CO}$ & $N_{\rm [CI]}/N_{\rm CO}$\\  \hline
Ring  &A& 49 & 0.23 &  0.12 & 1.2${+0.3 \atop -0.1}$ $\times$10$^{18}$ & 9.5${+5.3 \atop -3.0}$ $\times$10$^{18}$ & 1.8 $\pm$ 0.1 & 0.13${+0.10 \atop -0.05}$ \\  
        &B& 49 & 0.24 &  0.13 & 1.2${+0.5 \atop -0.1}$ $\times$10$^{18}$ & 1.0${+0.6 \atop -0.3}$ $\times$10$^{19}$ &  1.6 $\pm$ 0.1 &  0.12${+0.12 \atop -0.05}$ \\     
        &C& 49 & 0.16 &  0.09 & 9.4${+1.5 \atop -0.5}$ $\times$10$^{17}$ & 9.0${+5.3 \atop -3.0}$ $\times$10$^{18}$ &  1.4 $\pm$ 0.1 &  0.10${+0.08 \atop -0.04}$\\    
        &D& 49 & 0.20 &  0.09 & 9.1${+1.6 \atop -0.5}$ $\times$10$^{17}$ & 8.0${+4.6 \atop -2.6}$ $\times$10$^{18}$ &  1.6 $\pm$ 0.1 &  0.11${+0.08 \atop -0.05}$ \\   
        &E& 49 & 0.10  &  0.05 & 5.3 ${+0.7 \atop -0.5}$ $\times$10$^{17}$ & 4.2${+2.8 \atop -1.6}$ $\times$10$^{18}$ &  1.9 $\pm$ 0.2 &  0.13${+0.10 \atop -0.06}$\\    
        &F& 49 &  0.03 &  0.02 & 3.4${+1.3 \atop -1.1}$ $\times$10$^{16}$  & 9.0${+7.8 \atop -4.2}$ $\times$10$^{17}$ &  0.56 $\pm$ 0.2 &  0.04${+0.06 \atop -0.02}$ \\ \hline 
Finger&A& 25 & ---    &--- & --- & --- & --- & --- \\
         &B& 25 & ---    & --- & --- & --- & --- &  --- \\  
         &C& 25 & 0.13 & 0.03 & 3.0${+0.9 \atop -0.6}$ $\times$10$^{16}$ & 4.9${+2.3 \atop -1.9}$ $\times$10$^{16}$  & 5.3 $\pm$ 2.0  & 0.61 ${+0.69 \atop -0.28}$\\ 
         &D& 25 & 0.07 & ---   & 2.6${+1.0 \atop -0.8}$ $\times$10$^{16}$ & ---  & --- &---\\ 
         &E& 25  & 0.27 & 0.07 & 1.6${+0.4 \atop -0.2}$ $\times$10$^{17}$  & 4.5${+1.3 \atop -1.1}$ $\times$10$^{17}$   & 2.7 $\pm$ 0.4 &  0.36${+0.23 \atop -0.11}$\\ 
         &F& 25  & 0.31 & 0.08  & 1.5${+0.4 \atop -0.2}$ $\times$10$^{17}$  & 7.6${+2.0 \atop -1.7}$ $\times$10$^{17}$  & 1.6 $\pm$ 0.2 & 0.20${+0.12 \atop -0.06}$\\  
\hline
\end{tabular}}\label{tbl_prop}
\begin{tabnote}
\footnotemark[$*$] $\overline{\tau}$ denotes averaged optical depth.
\footnotemark[$\dagger$] Errors of column densities are derived by 
varying the assumed $T_{\rm ex}$ of the Ring and Finger clouds from 30 K to 80 K and from 20 K to 30 K, respectively.
\end{tabnote}
\end{table*}
%==================table==================%

\begin{figure*}
 \begin{center}
  \includegraphics[width=15cm]{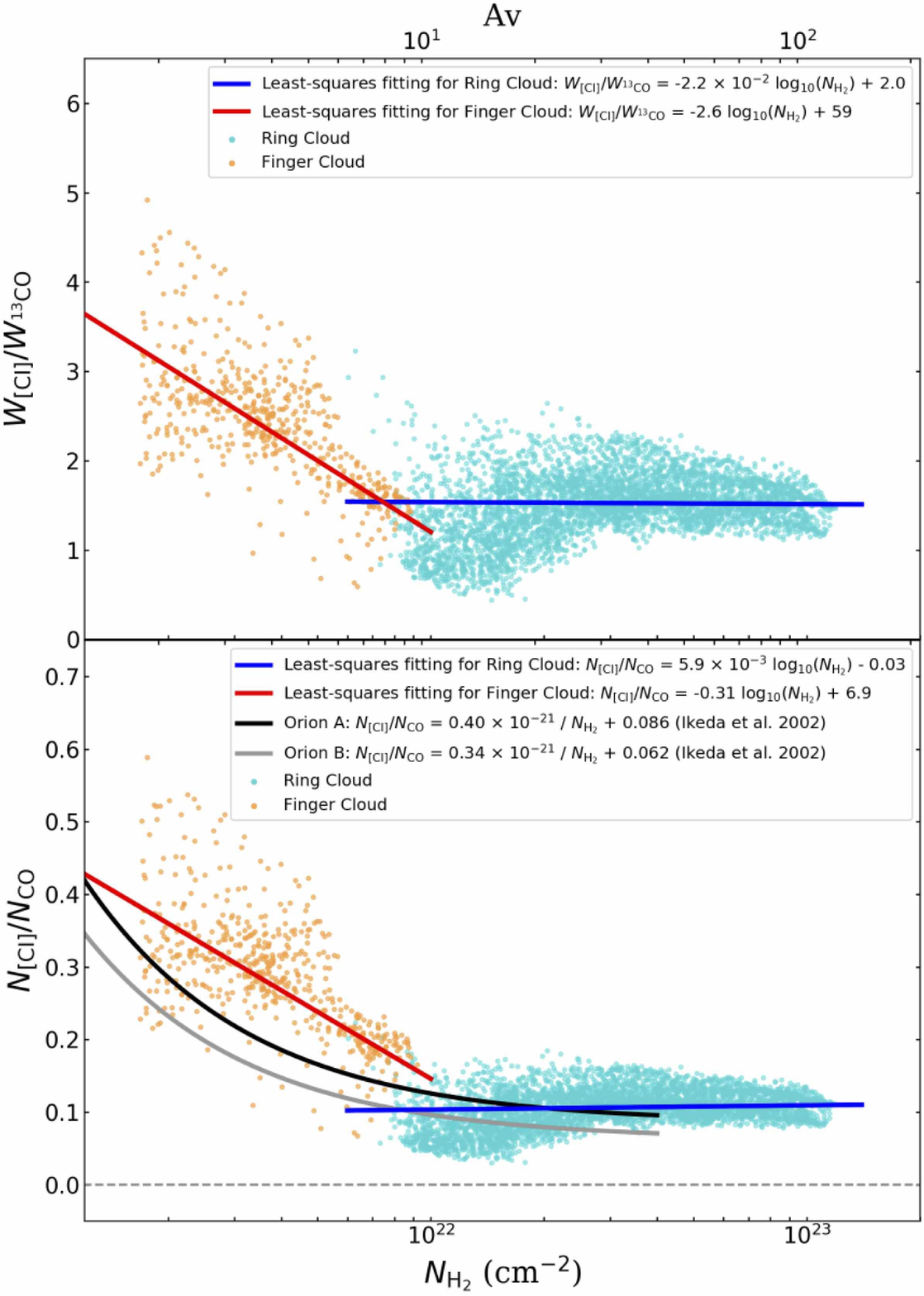} 
 \end{center}
\caption{
Point-by-point correlations between the ratios of [CI]/$^{13}$CO integrated intensity (Top), [CI]/CO column density (Bottom), and H$_2$ column density (Cyan: Ring cloud, Orange: Finger cloud).
Only pixels having intensities larger than the 3$\sigma$ noise levels in the data are plotted.
The 1$\sigma$ noise levels of the [CI] column density, CO column density, H$_2$ column density, [CI] intensity, and $^{13}$CO intensity of the Ring cloud are
1.4 $\times$ 10$^{16}$ cm$^{-2}$, 1.2 $\times$ 10$^{17}$ cm$^{-2}$, 1.2 $\times$ 10$^{21}$ cm$^{-2}$, 1.0 K km s$^{-1}$, and 0.5 K km s$^{-1}$, respectively.
The 1$\sigma$ noise levels of the [CI] column density, CO column density, H$_2$ column density, [CI] intensity, and $^{13}$CO intensity of the Finger cloud are
1.2 $\times$ 10$^{16}$ cm$^{-2}$, 5.6 $\times$ 10$^{16}$ cm$^{-2}$, 5.6 $\times$ 10$^{20}$ cm$^{-2}$, 0.9 K km s$^{-1}$, and 0.4 K km s$^{-1}$, respectively.
The blue and red lines in the top panel show the results of least-squares fitting for the Ring cloud and the Finger cloud: $R_{\rm{[CI]}/^{13}CO}$ = -2.2 ($\pm$ 1.6) $\times$ 10$^{-2}$ log$_{10}$($N_{\rm{H_2}}$) + 2.0
and $R_{\rm{[CI]}/^{13}CO}$ = -2.6 ($\pm$ 0.1) log$_{10}$($N_{\rm{H_2}}$) + 59 ($\pm$ 2), respectively.
The blue and red lines in the bottom panel show the results of least-squares fitting for the Ring cloud and the Finger cloud:
$N_{\rm{[CI]}}/N_{\rm{^{13}CO}}$ = 5.9 ($\pm$ 1.2) $\times$ 10$^{-3}$ log$_{10}$($N_{\rm{H_2}}$) - 0.03 ($\pm$ 0.03)
and $N_{\rm{[CI]}}/N_{\rm{^{13}CO}}$ = -0.47 ($\pm$ 0.02) log$_{10}$($N_{\rm{H_2}}$) + 10 ($\pm$ 0.4), respectively.
The black and gray lines in the bottom panel show the results of leat-squares fitting for Orion A and B clouds, respectively \citep{Ikeda2002}.
}
\label{CICO_H2column}
\end{figure*}

%==================table==================%
\begin{table*}
\tbl{Average and standard deviation values of [CI]/CO ratio}{%
\begin{tabular}{ccccc}
\hline  \hline
Cloud & \multicolumn{2}{c}{$W_{\rm [CI]}$/$W_{\rm ^{13}CO}$} & \multicolumn{2}{c}{$N_{\rm [CI]}$/$N_{\rm CO}$} \\ 
          & average  &   standard deviation                                             & average        & standard deviation                               \\ \hline
Ring  & 1.5 & 0.35 & 0.11 & 0.026           \\
Fing   & 2.4 & 0.69 & 0.29 & 0.084           \\
\hline
\end{tabular}}\label{tbl_std}
\begin{tabnote}
\end{tabnote}
\end{table*}
%==================table==================%

\begin{figure*}
 \begin{center}
  \includegraphics[width=16cm]{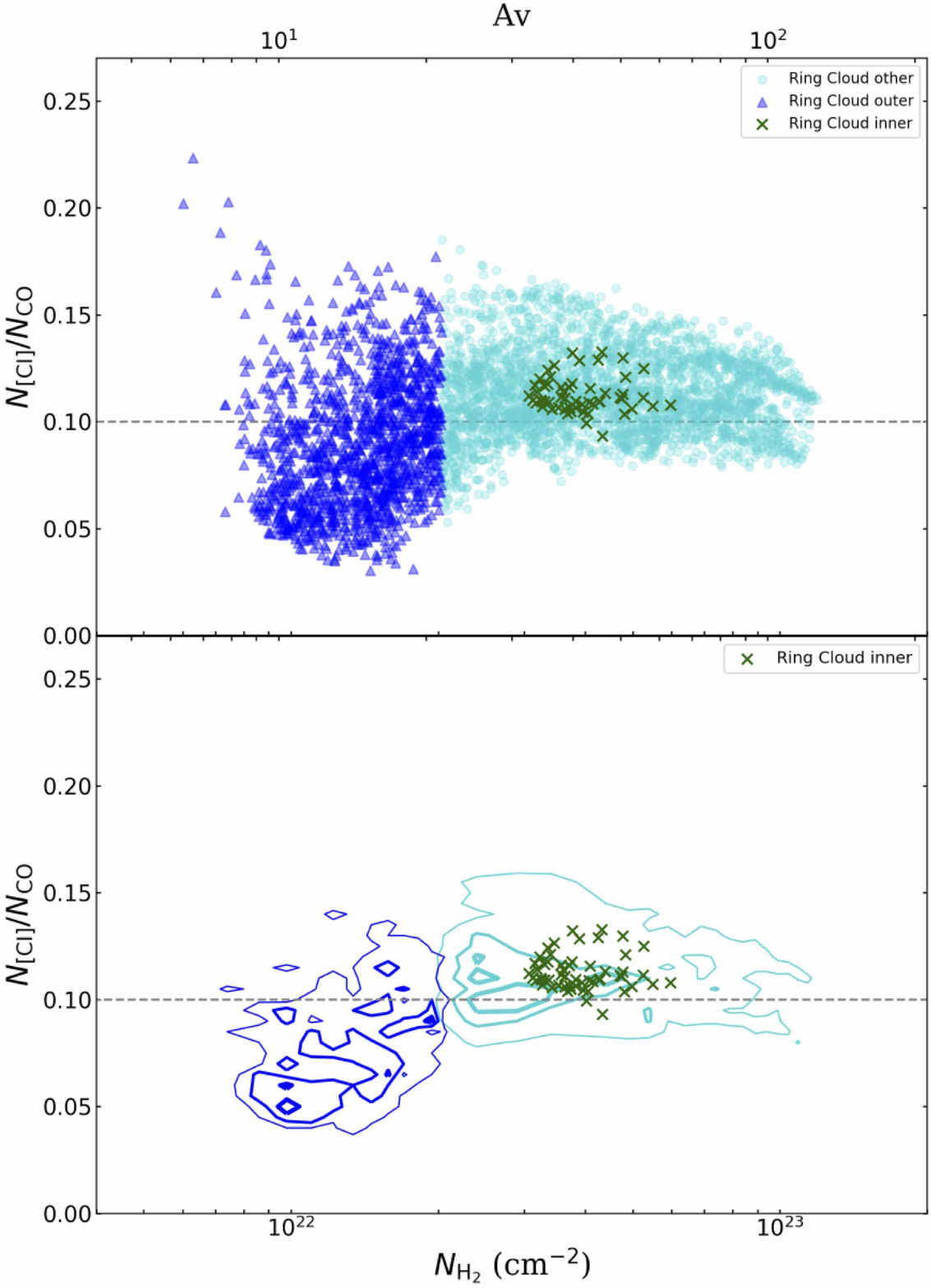} 
 \end{center}
\caption{Point-by-point correlations between the ratio of [CI]/CO column density and H$_2$ column density for the Ring cloud.
Top panel: blue, green, and cyan points show the outer, inner, and the other regions in the Ring cloud, respectively.
Bottom panel: cyan and blue contours show the distributions of all pixels for the other region and the outer region in the Ring cloud, respectively
(Other region: 6, 18, and 30 independent data points per 2.5 $\times$ 10$^{19}$ ((5.0 $\times$ 10$^{21}$) $\times$ 0.005) cell,
Outer region: 4, 9, and 14 independent data points per 0.6 $\times$ 10$^{19}$ ((1.2 $\times$ 10$^{21}$) $\times$ 0.005) cell).
}
\label{Ring_in_out}
\end{figure*}

\begin{figure*}
 \begin{center}
  \includegraphics[width=16cm]{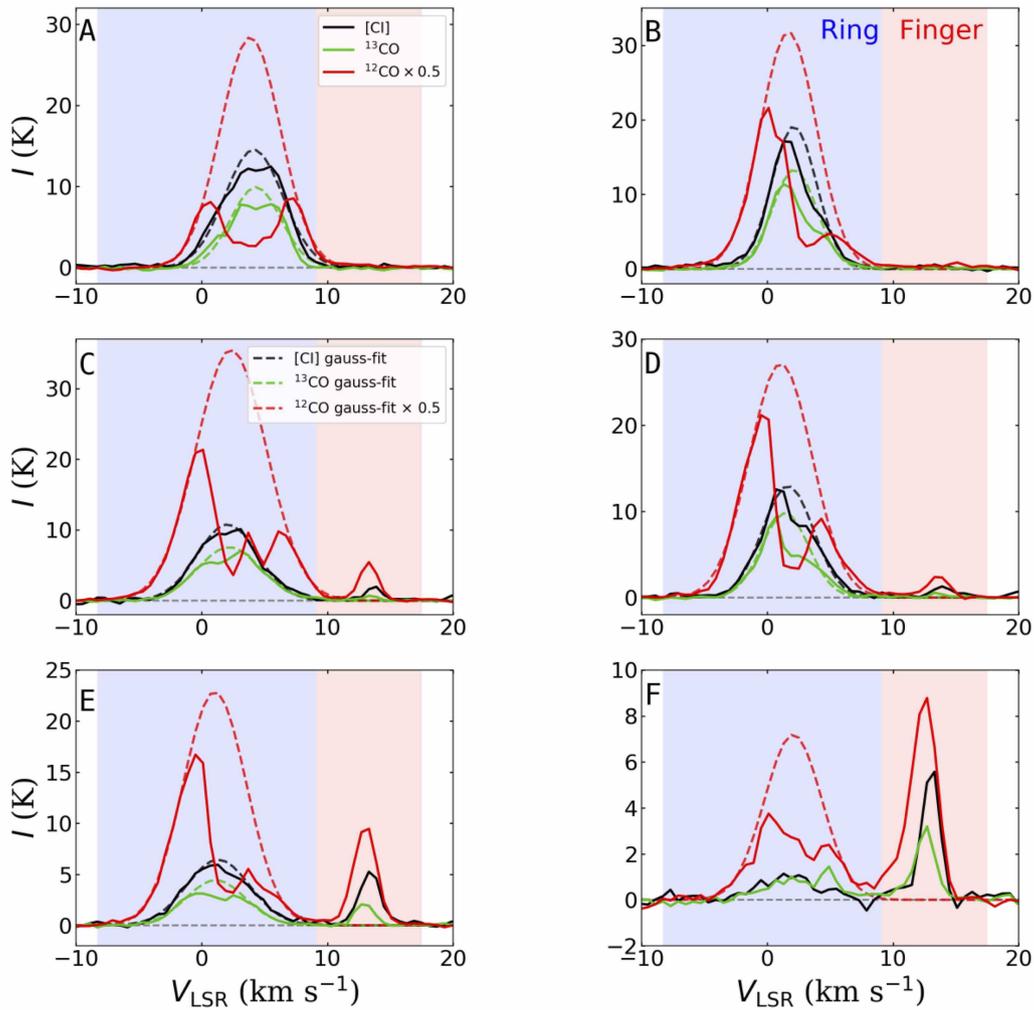} 
 \end{center}
\caption{\
Detected spectra (solid lines, same as in the Figure \ref{spectrum}) and gaussian-fitted spectra (dashed lines, only for the Ring cloud) of the [CI], $^{13}$CO, and $^{12}$CO emissions for the six [CI] peaks.
The blue and red areas indicate the velocity range of the Ring cloud (-8.3--9.1 km s$^{-1}$) and the Finger cloud (9.1--17.5 km s$^{-1}$), respectively.
We did not fit gaussian function to [CI] and $^{13}$CO spectra of peak F because we could not detect any self-absorption feature of [CI] and $^{13}$CO emission in the peak.
}
\label{Gauss_fit}
\end{figure*}
\clearpage

\begin{ack}
We are grateful to the ASTE staff for operating the ASTE and helping us with data reduction.
We would like to thank Hauyu Baobab Liu for helpful discussions.
We also would like to thank the anonymous referee for a carful reading and thoughtful suggestions that significantly improved this paper.
This work is supported by JSPS KAKENHI Grant Nos. 15H02074 and 18H05441.
\end{ack}

{}

\clearpage
\appendix
\section{Velocity-channel distributions of RCW38}\label{sec:apd_1}
\begin{figure*}
 \begin{center}
  \includegraphics[width=16cm]{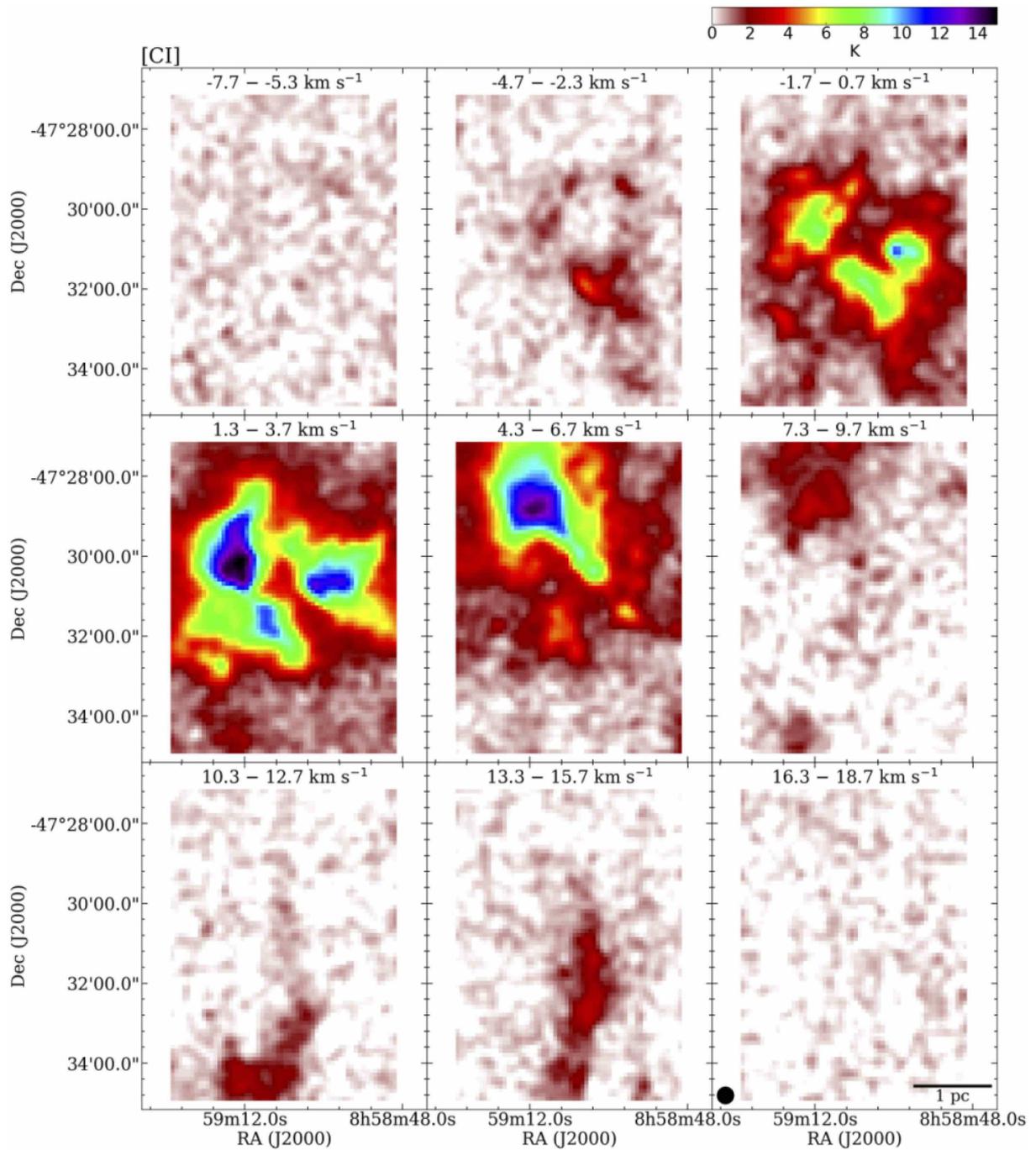} 
 \end{center}
\caption{[CI] velocity-channel map of RCW38.
These data were smoothed to an HPBW of 26$^{\prime \prime}$ with a 2D Gaussian function, and
the rms noise level was reduced to a typical value of 0.5 K in $T_{\rm MB}$.
The black filled circle at the lower left corner in the bottom right panel shows the HPBW size.
}
\label{channel_ci}
\end{figure*}

\begin{figure*}
 \begin{center}
  \includegraphics[width=16cm]{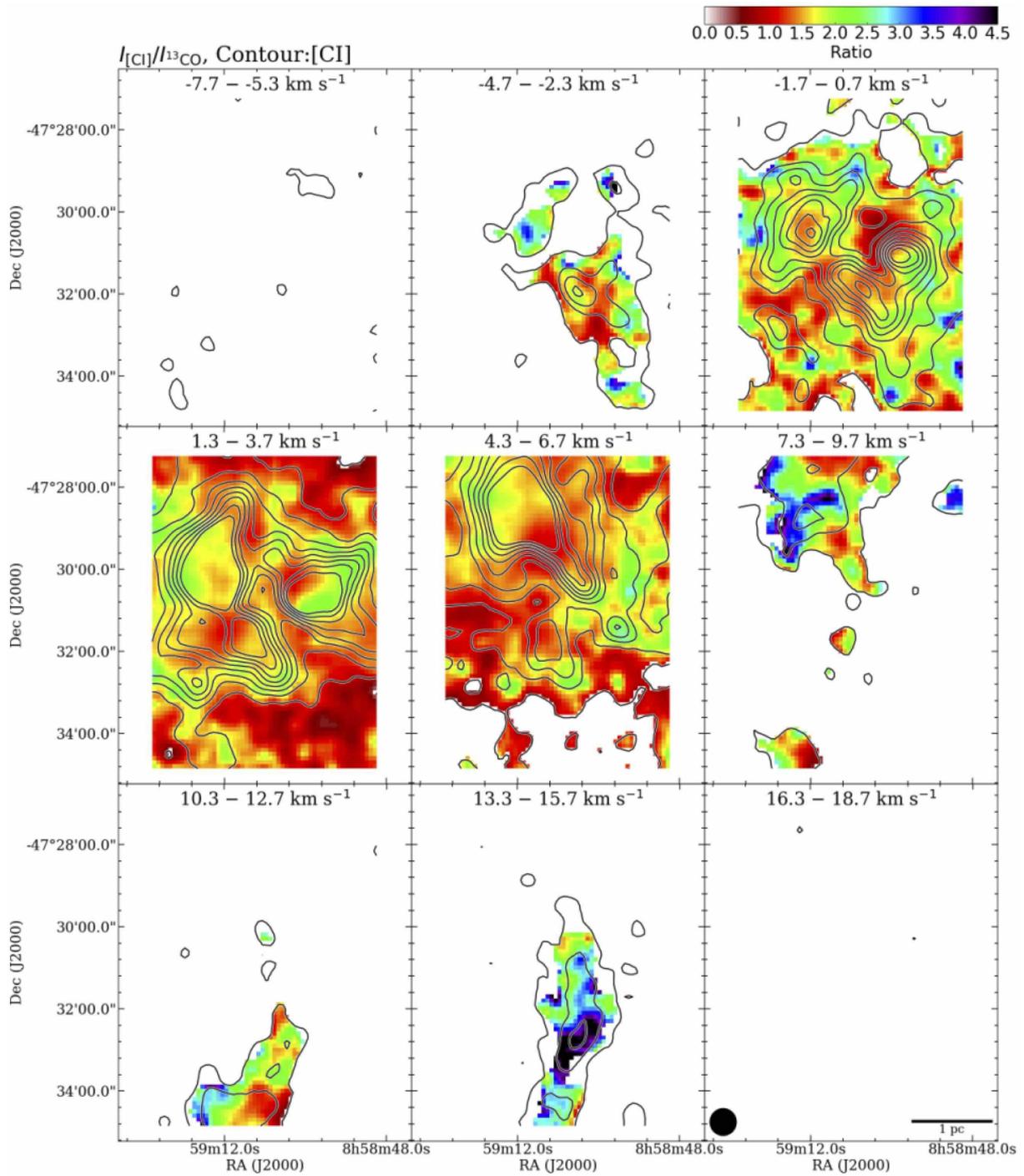} 
 \end{center}
\caption{Ratio of [CI] / $^{13}$CO line velocity-channel map of RCW38.
The black contours show the [CI] distribution (Contour levels: 3$\sigma$, 5$\sigma$, 13$\sigma$, 18$\sigma$, 23$\sigma$, 28$\sigma$, 33$\sigma$, 38$\sigma$, and 43$\sigma$, 1$\sigma$ = 0.2 K).
The black filled circle at the lower left corner in the bottom right panel shows the HPBW of the [CI] and CO data (40$^{\prime \prime}$).
}
\label{channel_cicoratio}
\end{figure*}

Velocity-channel distributions of RCW38 are shown in this appendix.
Figure \ref{channel_ci} and Figure \ref{channel_cicoratio} show the velocity-channel map of [CI] emission and [CI] / $^{13}$CO intensity ratio, respectively, for every 2.4 km s$^{-1}$.

\clearpage
\section{Results using $T_{\rm ex}$ derived from peak intensity of the $^{12}$CO(1-0)} \label{sec:apd_2}
In Sections \ref{sec:res} and \ref{sec:dis}, we presented the results in which $T_{\rm ex}$ was assumed to be uniform in both the Ring and the Finger clouds.
To assess the possible effects by variation of $T_{\rm ex}$, we present the results of [CI] parameters by employing a non-uniform distribution of $T_{\rm ex}$.
The results confirm that the present results of the [CI]/CO abundance ratio are robust.
To note, we did not calculate the column densities in the region with the ratio of $^{13}$CO(1-0)/$^{12}$CO(1-0) peak intensity of more than 1/3,
which corresponds to $\tau_{\rm ^{13}CO}$ of more than $\sim$ 0.4 estimated from the following equation:
\begin{equation}
\frac{T_{\rm MB} (\rm ^{12}CO)}{T_{\rm MB} (\rm ^{13}CO)} = \frac{1 - e^{-\tau_{\rm ^{12}CO}}}{1 - e^{-\tau_{\rm ^{13}CO}}} .
\end{equation}
and from the assumption that the abundance ratio of $^{12}$CO/$^{13}$CO is 77 \citep{Wilson1994}.
This is because the region is considered to be strongly affected by self-absorption (uncertainty of $T_{\rm ex}$ in this region is considered to be very large).

\begin{figure*}
 \begin{center}
  \includegraphics[width=16cm]{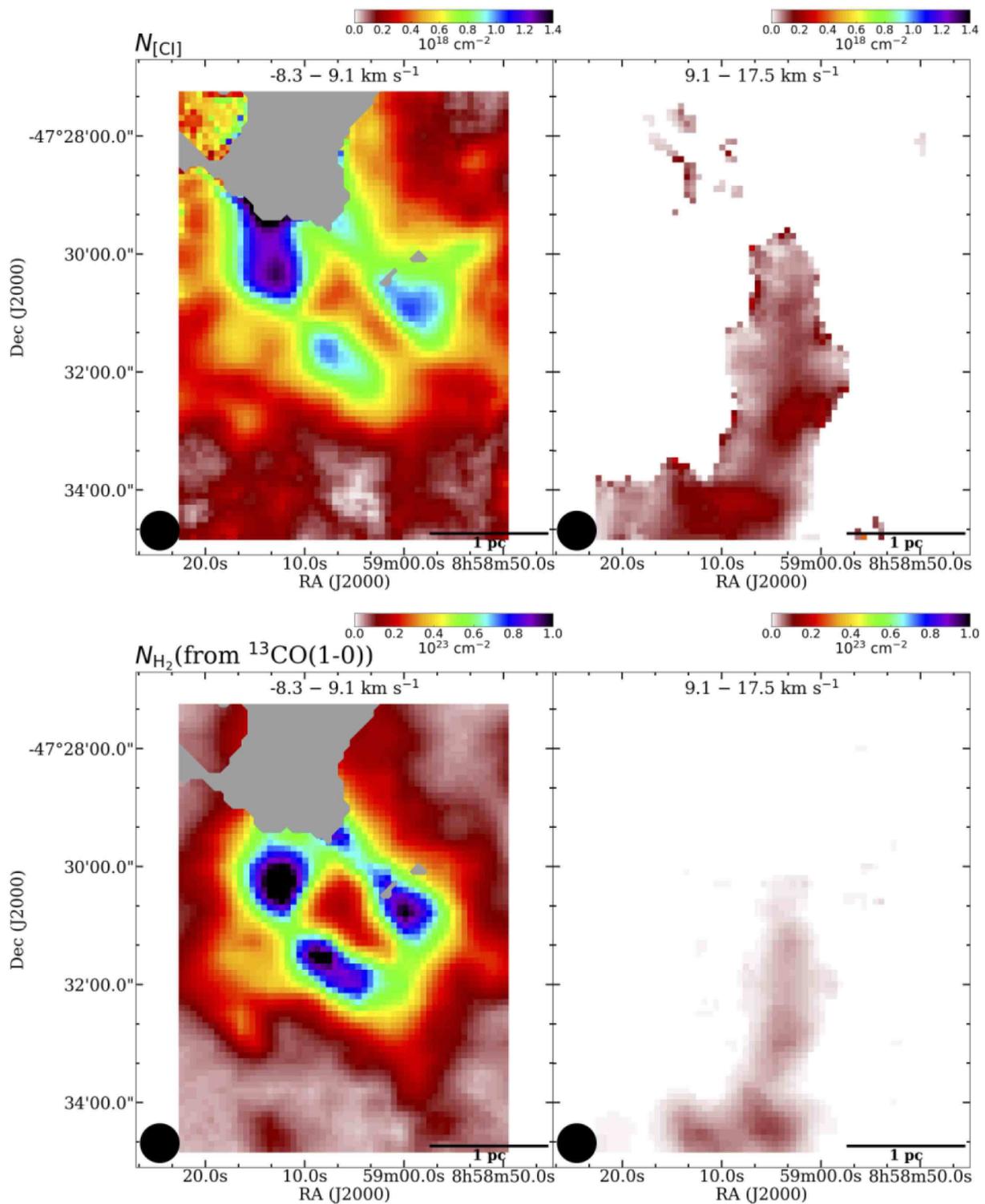} 
 \end{center}
\caption{
[CI] (Top) and H$_2$ (Bottom) column density distributions of RCW38 (Left: Ring cloud, Right: Finger cloud) using $T_{\rm ex}$ derived from the peak intensity of the $^{12}$CO(1-0).
The gray regions in the Ring cloud indicate the masked region derived from the ratio of $^{13}$CO(1-0) / $^{12}$CO(1-0) peak intensity.
The black filled circles at the lower left corners show the HPBW of the [CI] and CO data (40$^{\prime \prime}$).
}
\label{CIH2_column_pTex}
\end{figure*}

\begin{figure*}
 \begin{center}
  \includegraphics[width=16cm]{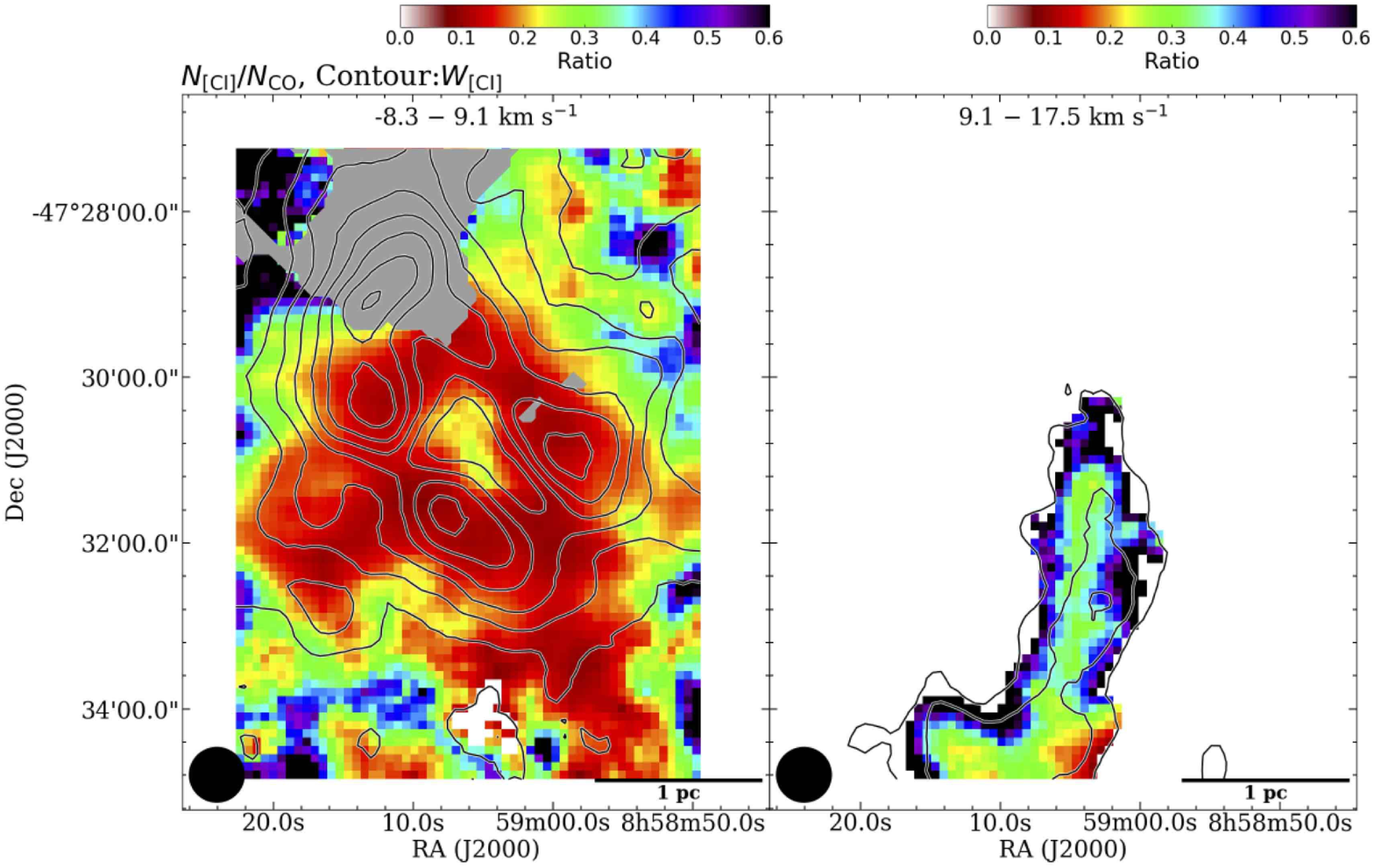}
 \end{center}
\caption{
Ratio of [CI]/CO column density ($N_{\rm [CI]}/N_{\rm CO}$) distribution of RCW38 (Left: Ring cloud, Right: Finger cloud) using $T_{\rm ex}$ derived from the peak intensity of the $^{12}$CO(1-0).
The black contours show the integrated intensity distribution of [CI] emission.
Contour levels of the Ring cloud are 3$\sigma$, 13$\sigma$, 23$\sigma$, 33$\sigma$, 43$\sigma$, 53$\sigma$, 63$\sigma$, 73$\sigma$, and 83$\sigma$ (1$\sigma$ = 1.0 km s$^{-1}$). 
Contour levels of the Finger cloud are 3$\sigma$, 8$\sigma$, and 13$\sigma$ (1$\sigma$ = 0.9 km s$^{-1}$).
The gray regions in the Ring cloud indicate the masked region derived from the ratio of $^{13}$CO(1-0) / $^{12}$CO(1-0) peak intensity.
The black filled circles at the lower left corners show the HPBW of the [CI] and CO data (40$^{\prime \prime}$).
}
\label{CICO_column_pTex}
\end{figure*}

\begin{figure*}
 \begin{center}
  \includegraphics[width=16cm]{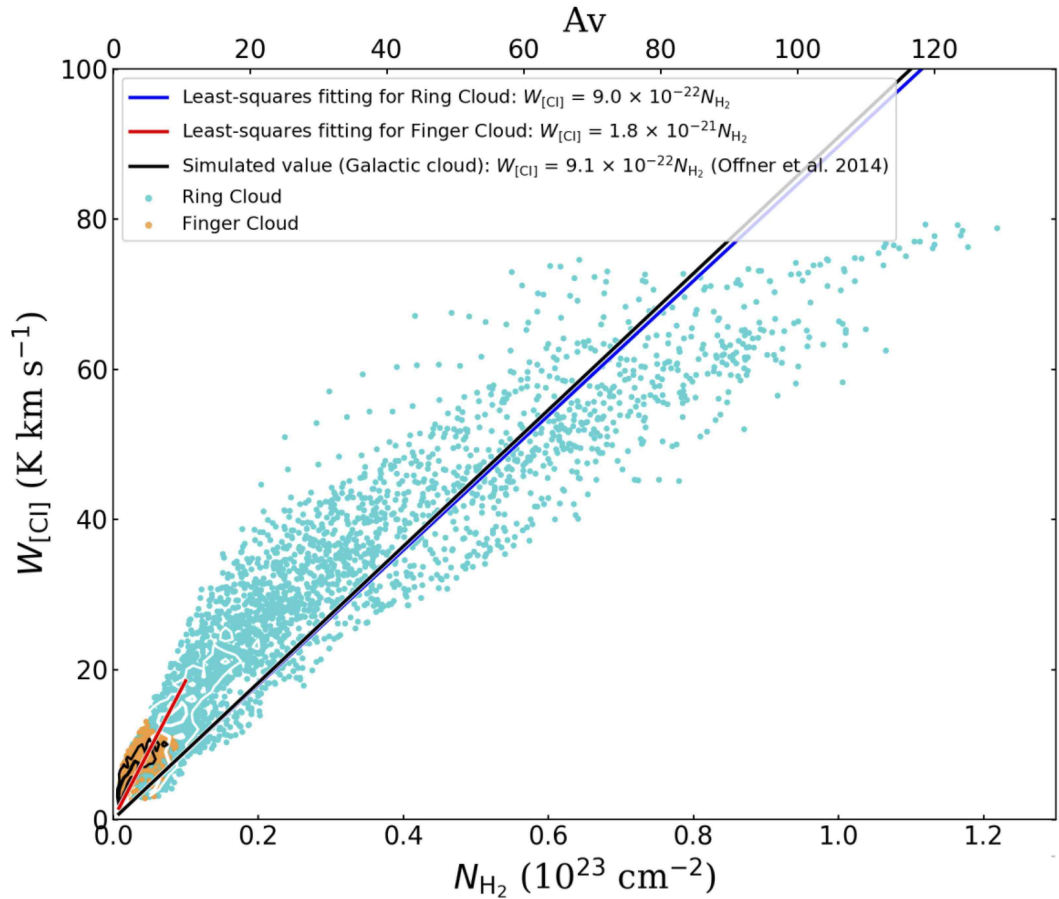} 
 \end{center}
\caption{
Point-by-point correlations between the integrated intensity of [CI] emission and H$_2$ column density using $T_{\rm ex}$ derived from the peak intensity of the $^{12}$CO(1-0) (Cyan: Ring cloud, Orange: Finger cloud).
Only pixels having intensities larger than the 3$\sigma$ noise levels in the data are plotted.
The 1$\sigma$ noise level of the [CI] integrated intensity and H$_2$ column density of the Ring cloud are 1.0 K km s$^{-1}$ and 7.0 $\times$ 10$^{20}$ cm$^{-2}$, respectively,
and those of the Finger cloud are 0.9 K km s$^{-1}$ and 3.0 $\times$ 10$^{20}$ cm$^{-2}$, respectively.
The blue and red lines show the results of the least-squares fitting for the Ring cloud and the Finger cloud:
$W_{\rm [CI]}$ = 9.0 ($\pm$ 0.04) $\times$ 10$^{-22}$ $N_{\rm{H_2}}$
and $W_{\rm [CI]}$ = 1.8 ($\pm$ 0.03) $\times$ 10$^{-21}$ $N_{\rm{H_2}}$, respectively.
The black line shows the simulated value for a typical Galactic cloud: $W_{\rm [CI]}$ = 9.1 $\times$ 10$^{-22}$ $N_{\rm{H_2}}$ \citep{Offner2014}.
The white and black contours show the distributions of all pixels for the Ring cloud and the Finger cloud, respectively
(Ring cloud: 10 and 20 independent data points per 1.5 $\times$ 10$^{21}$ cell, Finger cloud: 5 and 15 independent data points per 0.5 $\times$ 10$^{21}$ cell).
}
\label{CIint-H2column_pTex}
\end{figure*}

Figure \ref{CIH2_column_pTex} shows the [CI] (top) and H$_2$ (bottom) column density distributions of RCW38.
The [CI] column densities of the Ring and Finger clouds are (0.1--1.3) $\times$ 10$^{18}$ and (0.1--0.2) $\times$ 10$^{18}$ cm$^{-2}$, respectively.
The H$_2$ column densities of the Ring and Finger clouds are (0.1--1.0) $\times$ 10$^{23}$ and $\sim$ 0.1 $\times$ 10$^{23}$ cm$^{-2}$, respectively.
Figure \ref{CICO_column_pTex} shows the distribution of the ratio of [CI]/CO column density ($N_{\rm [CI]}$/$N_{\rm CO}$) of RCW38.
The ratio in both the Ring and Finger clouds are 0.1--0.6.
Figure \ref{CIint-H2column_pTex} shows the relationship between [CI] integrated intensity ($W_{\rm [CI]}$) and H$_2$ column density ($N_{\rm H_2}$).
The conversion factors from $W_{\rm [CI]}$ to $N_{\rm H_2}$ ($X_{\rm [CI]}$ = $N_{\rm H_2}$/$W_{\rm [CI]}$) for the Ring and Finger clouds are
$X_{\rm [CI]}$ = 1.1 ($\pm$ 0.005) $\times$ 10$^{21}$ cm$^{-2}$ K$^{-1}$ km$^{-1}$ s
(corresponding to $W_{\rm [CI]}$ = 9.0 ($\pm$0.04) $\times$ 10$^{-22}$ $N_{\rm H_2}$) and
$X_{\rm [CI]}$ = 5.6 ($\pm$ 0.09) $\times$ 10$^{20}$ cm$^{-2}$ K$^{-1}$ km$^{-1}$ s
(corresponding to $W_{\rm [CI]}$ = 1.8 ($\pm$0.03) $\times$ 10$^{-22}$ $N_{\rm H_2}$), respectively.

\begin{figure*}
 \begin{center}
  \includegraphics[width=15cm]{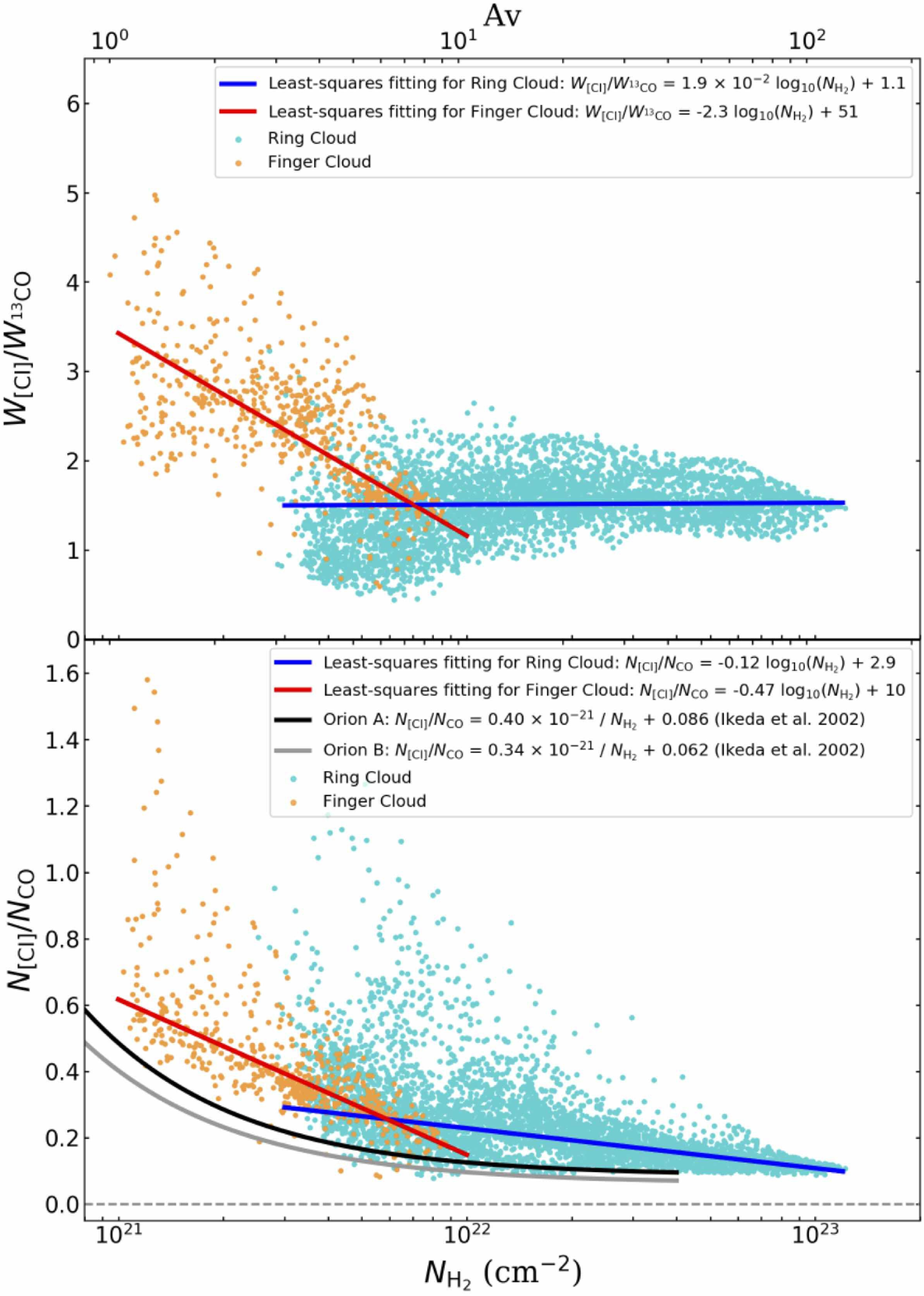} 
 \end{center}
\caption{
Point-by-point correlations between the ratio of [CI]/$^{13}$CO integrated intensity (Top), [CI]/CO column density (Bottom), and H$_2$ column density (Cyan: Ring cloud, Orange: Finger cloud)
using $T_{\rm ex}$ derived from the peak intensity of the $^{12}$CO(1-0).
Only pixels having intensities larger than the 3$\sigma$ noise levels in the data are plotted.
The 1$\sigma$ noise levels of the [CI] column density, CO column density, H$_2$ column density, [CI] intensity, and $^{13}$CO intensity of the Ring cloud are
1.8 $\times$ 10$^{16}$ cm$^{-2}$, 7.0 $\times$ 10$^{16}$ cm$^{-2}$, 7.0 $\times$ 10$^{20}$ cm$^{-2}$, 1.0 K km s$^{-1}$, and 0.5 K km s$^{-1}$, respectively.
The 1 $\sigma$ noise levels of the [CI] column density, CO column density, H$_2$ column density, [CI] intensity, and $^{13}$CO intensity of the Finger cloud are
1.2 $\times$ 10$^{16}$ cm$^{-2}$, 3.0 $\times$ 10$^{16}$ cm$^{-2}$, 3.0  $\times$ 10$^{20}$ cm$^{-2}$, 0.9 K km s$^{-1}$, and 0.4 K km s$^{-1}$, respectively.
The blue and red lines in the top panel show the results of the least-squares fitting for the Ring cloud and the Finger cloud:
$R_{\rm{[CI]}/^{13}CO}$ = 1.9 ($\pm$ 1.2) $\times$ 10$^{-2}$ log$_{10}$($N_{\rm{H_2}}$) + 1.1 ($\pm$ 0.3)
and $R_{\rm{[CI]}/^{13}CO}$ = -2.3 ($\pm$ 0.1) log$_{10}$($N_{\rm{H_2}}$) + 51 ($\pm$ 2), respectively.
The blue and red lines in the bottom panel show the results of the least-squares fitting for the Ring cloud and the Finger cloud:
$N_{\rm{[CI]}}/N_{\rm{^{13}CO}}$ = -0.12 ($\pm$ 0.002) log$_{10}$($N_{\rm{H_2}}$) + 2.9 ($\pm$ 0.05)
and $N_{\rm{[CI]}}/N_{\rm{^{13}CO}}$ = -0.47 ($\pm$ 0.02) log$_{10}$($N_{\rm{H_2}}$) + 10 ($\pm$ 0.4), respectively.
The black and gray lines in the bottom panel show the results of leat-squares fitting for the Orion A and B clouds, respectively \citep{Ikeda2002}.
}
\label{CICO_H2column_pTex}
\end{figure*}

The top panel in Figure \ref{CICO_H2column_pTex} shows the relationship between $W_{\rm [CI]}/W_{\rm ^{13}CO}$ and $N_{\rm H_2}$,
and the bottom panel in Figure \ref{CICO_H2column} shows the relationship between $N_{\rm [CI]}/N_{\rm CO}$ and $N_{\rm H_2}$.
In the low-$A_V$ region ($A_V$ $\le$ 10 mag), $W_{\rm [CI]}/W_{\rm ^{13}CO}$ decreases with $A_V$ from $\sim$ 5 to $\sim$ 1.
In the high-$A_V$ region ($A_V$ $>$ 10 mag), $W_{\rm [CI]}/W_{\rm ^{13}CO}$ is almost constant ($\sim$ 1.5) for $A_V$ of up to 100 mag.
A similar trend is seen in the relationship between $N_{\rm [CI]}/N_{\rm CO}$ and $A_V$ ($N_{\rm H_2}$).
In the low-$A_V$ region ($A_V$ $\le$ 10 mag), $N_{\rm [CI]}/N_{\rm CO}$ decreases with $A_V$ from $\sim$ 1.6 to $\sim$ 0.2.
In the high-$A_V$ region ($A_V$ $>$ 10 mag), $N_{\rm [CI]}/N_{\rm CO}$ gradually decreases with $A_V$ from $\sim$ 1.2 to $\sim$ 0.1
and $N_{\rm [CI]}/N_{\rm CO}$ is $\sim$ 0.1 for $A_V$ of up to 100 mag.

\begin{figure*}
 \begin{center}
  \includegraphics[width=16cm]{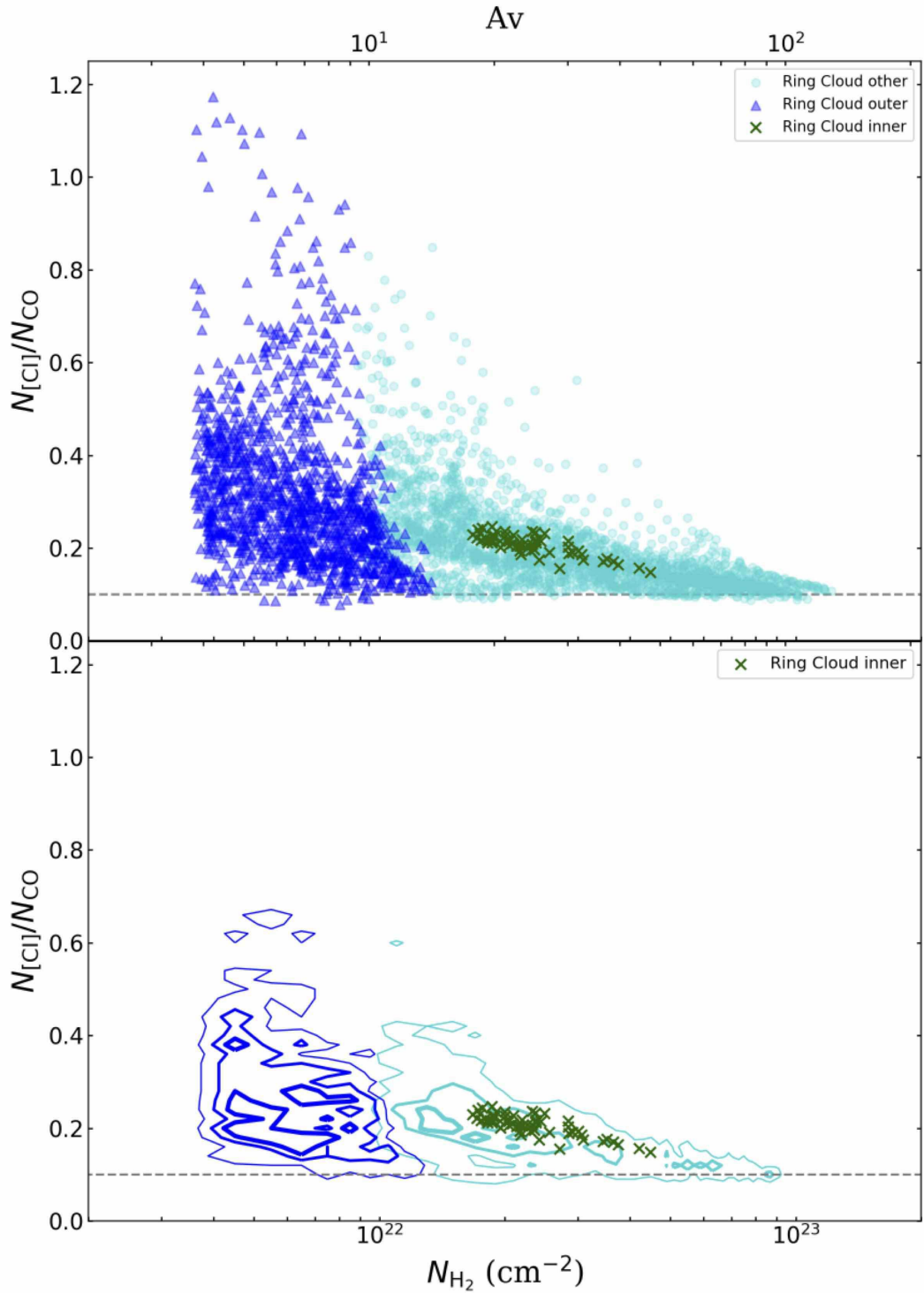} 
 \end{center}
\caption{Point-by-point correlations between the ratio of [CI]/CO column density and the H$_2$ column density for the Ring cloud using $T_{\rm ex}$ derived from the peak intensity of the $^{12}$CO(1-0).
Top panel: blue, green, and cyan points show the outer, inner, and the other regions in the Ring cloud, respectively.
Bottom panel: cyan and blue contours show the distributions of all pixels for the other region and the outer region in the Ring cloud, respectively
(Other region: 4, 12, and 20 independent data points per 4.0 $\times$ 10$^{19}$ ((2.0 $\times$ 10$^{21}$) $\times$ 0.02) cell,
Outer region: 3, 8, and 13 independent data points per 2.0 $\times$ 10$^{19}$ ((1.0 $\times$ 10$^{21}$) $\times$ 0.02) cell).
}
\label{Ring_in_out_pTex}
\end{figure*}

Figure \ref{Ring_in_out_pTex} shows the relationship between $N_{\rm [CI]}/N_{\rm CO}$ and $N_{\rm H_2}$ for the inner, outer, and other regions of the Ring cloud.
The ratio of $N_{\rm [CI]}/N_{\rm CO}$ in the inner region is relatively larger than that in the other region.
The ratio of $N_{\rm [CI]}/N_{\rm CO}$ in the outer region in slightly lower than that in the other region, especially for $N_{\rm H_2}$ of $\sim$ 10$^{22}$ cm$^{-2}$.

\end{document}